\documentclass[manuscript,screen]{acmart}
\AtBeginDocument{%
  \providecommand\BibTeX{{%
    \normalfont B\kern-0.5em{\scshape i\kern-0.25em b}\kern-0.8em\TeX}}}

\setcopyright{acmcopyright}
\copyrightyear{2024}
\acmYear{2024}
\acmDOI{XXXXXXX.XXXXXXX}

\usepackage[inline]{enumitem}
\usepackage{array}
\usepackage{multirow}
\usepackage{rotating}
\usepackage{tabularray}
\usepackage{color}
\usepackage{tikz}
\usetikzlibrary{calc}
\usepackage{subcaption}

\newcommand{\repourl}{https://github.com/meluso/open-source-credit-survey}

\makeatletter
  \if@ACM@anonymous
    \renewcommand{\repourl}{https://anonymous.4open.science/r/open-source-credit-survey-5AE6}
  \fi
\makeatother

\begin{document}

\title[Invisible Labor in OSS]{Invisible Labor in Open Source Software Ecosystems}

\author{John Meluso}
\email{john.meluso@cornell.edu}
\orcid{0000-0001-8200-8150}
\affiliation{%
  \institution{Cornell University}
  \city{Ithaca}
  \state{New York}
  \country{USA}
  \postcode{14850}
}
\affiliation{%
  \institution{University of Vermont}
  \streetaddress{33 Colchester Ave.}
  \city{Burlington}
  \state{Vermont}
  \country{USA}
  \postcode{05405}
}

\author{Amanda Casari}
\orcid{0000-0002-3201-8964}
\affiliation{%
  \institution{Google LLC}
  \city{Burlington}
  \state{Vermont}
  \country{USA}
}
\affiliation{%
  \institution{University of Vermont}
  \city{Burlington}
  \state{Vermont}
  \country{USA}
  \postcode{05405}
}
\email{amcasari@google.com}

\author{Katie McLaughlin}
\orcid{0000-0003-2684-4220}
\affiliation{%
  \institution{Google LLC}
  \city{Melbourne}
  \state{Victoria}
  \country{Australia}
}
\email{glasnt@google.com}

\author{Milo Z. Trujillo}
\orcid{0000-0003-0143-5869}
\affiliation{%
  \institution{Northeastern University}
  \city{Boston}
  \state{Massachusetts}
  \country{USA}
  \postcode{02115}
}
\affiliation{%
  \institution{University of Vermont}
  \city{Burlington}
  \state{Vermont}
  \country{USA}
  \postcode{05405}
}
\email{milo.trujillo@uvm.edu}

\renewcommand{\shortauthors}{Meluso, et al.}

\begin{abstract}
Invisible labor is work that is either not fully visible or not appropriately compensated.
In open source software (OSS) ecosystems, essential tasks that do not involve code (like content moderation) often become invisible to the detriment of individuals and organizations.
However, invisible labor is sufficiently difficult to measure that we do not know how much of OSS activities are invisible.
Our study addresses this challenge, demonstrating that roughly half of OSS work is invisible.
We do this by developing a cognitive anchoring survey technique that measures OSS developer self-assessments of labor visibility and attribution.
Survey respondents ($n=142$) reported that their work is more likely to be invisible (2 in 3 tasks) than visible, and that half (50.1\%) is uncompensated.
Priming participants with the idea of visibility caused participants to think their work was more visible, and that visibility was less important, than those primed with invisibility.
We also found evidence that tensions between attribution motivations probably increase how common invisible labor is.
This suggests that advertising OSS activities as ``open'' may lead contributors to overestimate how visible their labor actually is.
Our findings suggest benefits to working with varied stakeholders to make select, collectively valued activities visible, and increasing compensation in valued forms (like attribution, opportunities, or pay) when possible.
This could improve fairness in software development while providing greater transparency into work designs that help organizations and communities achieve their goals.
\end{abstract}




\maketitle

%%%%%%%%%%%%%%%%%%%%%%%%%%%%%%%%%%%%%%%%%%%%%%%%%%%%%%%%%%%%%%%%%%%%%%%%%%%%%%%%%%%%%%%%%%%%%%%%%%%
\section{Introduction}
\label{sec:introduction}

\begin{figure}[b]
  \centering
  \includegraphics[width=\linewidth]{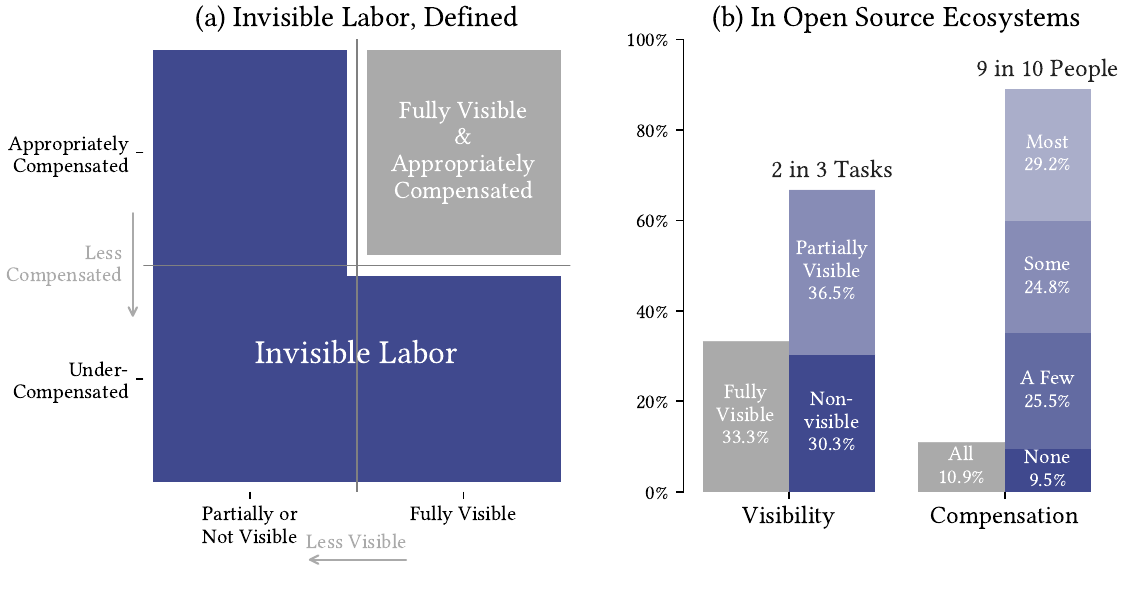}
  \caption{(a) Invisible labor is labor that is not fully visible and/or undercompensated. Labor is not fully visible when labor data doesn't exist, isn't shared, or isn't accessible. Labor is undercompensated when it doesn't receive enough credit, pay, or new opportunities, among others forms of compensation. (b) Our study demonstrates that roughly two thirds of labor in OSS is not visible, either by not being visible or only shared with one other person. We also found that about half of OSS labor activities do not receive credit, as demonstrated here by the 9 in 10 individuals who do not receive credit from some fraction of projects they contribute to.}
  \Description{(a) Invisible labor becomes so when it is not visible (because it doesn't exist, isn't shared, or isn't accessible) and/or when labor is undercompensated (because it doesn't receive credit, doesn't receive pay, doesn't facilitate new opportunities, etc.). (b) One third of labor in our study was visible, one third partially visible, and one third invisible, and more than half of labor did not receive credit.}
  \label{fig:conceptual_overview}
\end{figure}

% Less visible tasks receive less credit than tasks that are easy to track

Creating and maintaining open source software (OSS) involve many diverse tasks beyond contributing code to software repositories \cite{Giuri2010Skills, Casari2021Open, Trinkenreich2020Hidden, Geiger2021Labor}. Tasks like conference organizing, community management, finance management, interaction design, data entry, authorship verification, and documenting have become essential to OSS production, repair, and safeguarding \cite{Gousios2008Measuring, DIgnazio2020Data, Ramin2020More, Young2021Which, Trinkenreich2020Hidden, Kalgutkar2019Code}. Acknowledging the breadth and interdependence of these activities, many OSS stakeholders now think about OSS in terms of ``ecosystems'' (encompassing related software repositories, dependencies, managing organizations, software users, other stakeholders, and their interactions) instead of the less comprehensive idea of ``projects'' (often limited to code and actions for independent repositories, exclusive of people and interdependencies) \cite{Germonprez2018Eight, Geiger2021Labor, Frluckaj2022Gender, Casari2023Repository, NSF2023POSE, Jergensen2011Onion, NSF2024SafeOSE}. Shifting our focus from projects to ecosystems --- from code to interdependent people and technologies --- is important because it reifies a long-held understanding in Computer-Supported Cooperative Work (CSCW) \cite{Suchman1995Making, Star1999Layers}: that much of the labor performed in support of software development is undocumented, unattributed to individuals, and ``practically invisible'' to outsiders and newcomers \cite{Young2021Which, Geiger2021Labor}. In other words, OSS ecosystems involve substantial \textit{invisible labor}: undercompensated or uncompensated activities that are ``rendered invisible...because they take place out of sight, or because they lack physical form altogether'' \cite{DIgnazio2020Data}.

% In turn, less credit often means less compensation for work and less organizational knowledge

Labor becomes invisible either when (1) labor data is not fully visible (because it does not exist, is not shared, or is not accessible) or when (2) labor is undercompensated (Fig. \ref{fig:conceptual_overview}a). While invisibility certainly offers some benefits (like anonymity, autonomy, process flexibility, and reduced coordination), it also creates significant challenges for individuals and organizations involved in OSS \cite{Suchman1995Making, Star1999Layers}. Lower task visibility reduces the extrinsic value of individuals' contributions \cite{Suchman1995Making, DIgnazio2020Data} because individuals receive less compensation, whether in terms of future opportunities \cite{Riehle2007Economic, Babcock2017Gender}, pay \cite{Toxtli2021Quantifying}, or social status \cite{Stewart2005Social}. Equally important, it risks delegitimizing less visible tasks and workers, making them more precarious \cite{Star1999Layers}. Consequently, individuals frequently burn out and stop performing crucial activities \cite{Geiger2021Labor} on which a growing number of scientists, organizations, and governments rely \cite{NSF2024SafeOSE}. Furthermore, these individuals are often members of minoritized groups \cite{Star1999Layers, Frluckaj2022Gender}, which reinforces demographic representation disparities \cite{Babcock2017Gender, DIgnazio2020Data}.

For organizations and governments, labor invisibility limits work transparency and therefore task measurement, coordination, and control \cite{Suchman1995Making}. This constrains organizations' knowledge into whom they fund in open source, what those individuals do, and how those activities shape valued outcomes \cite{Casari2021Open}. For instance, invisibility limits verifiability of authorship \cite{Kalgutkar2019Code} among other security challenges \cite{NSF2024SafeOSE}, increasingly important concerns in light of the 2024 XZ Utils scandal \cite{Greenberg2024Mystery}. In addition, computing researchers have long argued that invisibility prevents equity more than creates it despite the aforementioned benefits \cite{Suchman1995Making, Star1999Layers, DIgnazio2020Data}. Indeed, the persisting reliance of open source ecosystems on volunteerism necessitates creating communities where volunteers find benefit in performing all kinds of activities (i.e. not just coding) \cite{Fang2009Understanding, VonKrogh2012Carrots}.

% We don't know how much of OSS work is invisible

Not all work should --- or even can --- be visible. But given these limitations, some labor certainly should be more visible than it currently is. To make OSS labor (more) visible, we likely need to (a) recognize diverse types of work that individuals perform and (b) measure how much of those activities people perform. On the former, task taxonomies have seen modest success. For example, academics have started adopting contributorship taxonomies like CRediT to increase attribution and transparency \cite{Brand2015Authorship, Lariviere2016Contributorship, Holcombe2019Contributorship}. Likewise, software developers are experimenting with manual and automated attribution systems (e.g. All Contributors and octohatrack, respectively) with some success \cite{Young2021Which}.

Measuring amounts of invisible labor, though, has proven more elusive, particularly in OSS. Many automated measurement techniques (like GitHub's contribution activity visualizations) only measure \textit{code} as ``contributions,'' thereby conferring lower status to \textit{non-coding} activities. This provides limited, often overly reductive assessments of ``how much got done'' \cite{DIgnazio2020Data, Young2021Which}, can be less meaningful to workers than manual attribution \cite{Monroy-Hernandez2011Computers}, and confirms concerns of delegitimization. And while computational solutions hold potential for spaces like crowd work (e.g. work on crowdsourcing markets like Amazon Mechanical Turk) \cite{Toxtli2021Quantifying}, they seem unlikely to pass muster in open source ecosystems given the variety of computational and noncomputational interaction contexts involved (coding, conferences, direct chats, etc.). This leaves a need for research on both the prevalence of invisible labor in OSS ecosystems and factors influencing its prevalence. To address these gaps, this article answers two research questions:
\begin{enumerate}[label={\textbf{RQ\arabic*.}},leftmargin=2cm]
    \item How common is invisible labor in OSS ecosystems?
    \item What factors affect how commonly labor invisibility occurs in OSS ecosystems?
\end{enumerate}

We answer these questions by characterizing invisible labor in OSS ecosystems through an explanatory mixed-methods analysis \cite{Creswell2017Designing}, with three contributions. First, we introduce a survey technique that estimates labor visibility in OSS ecosystems through cognitive anchoring \cite{Kahneman1982Judgment, Tourangeau2018Survey}. Next, we demonstrate that roughly half of OSS activities are invisible (Fig. \ref{fig:conceptual_overview}b) by measuring OSS labor attribution and visibility across varied fields --- from scientific computing to app development. Third, we show that motivations and resource constraints often leave attribution practices at cross-purposes with one another, thereby limiting labor visibility and compensation.

% Article organization

This article begins by clearly defining labor visibility, invisible labor, and their roles in CSCW and OSS ecosystems (Sec. \ref{sec:background}). We then describe our methods (Sec. \ref{sec:methods}), including our participatory feminist approach, survey instrument design, and explanatory mixed-methods analyses. Next, we share results from our quantitative (Sec. \ref{sec:quantresults}) and qualitative (Sec. \ref{sec:qualresults}) analyses of a global sample of software developers. We close with a discussion (Sec. \ref{sec:discussion}) of what our findings mean for individuals and organizations involved in OSS along with design considerations for overcoming these challenges. Fig. \ref{fig:conceptual_overview} presents an overview of our findings.

%%%%%%%%%%%%%%%%%%%%%%%%%%%%%%%%%%%%%%%%%%%%%%%%%%%%%%%%%%%%%%%%%%%%%%%%%%%%%%%%%%%%%%%%%%%%%%%%%%%
\section{Background}
\label{sec:background}

To measure something that, by implication, one ``cannot see,'' it helps to understand what it means for work to be ``visible'' and ``invisible'' in the context of OSS. The following sections address these concepts in turn.

\subsection{Labor Visibility}
\label{subsec:background_laborVisibility}

When we call work ``visible,'' we typically do not mean that we can physically ``see'' the work as it happens. Instead, we use \textbf{\textit{visibility}} here as a metaphor that encompasses three qualities: how available information is, how widely shared information is, and how accessible that information is \cite{Stohl2016Managing, Poster2016Introduction}. If a piece of information is fully visible, that information has been recorded, shared widely, and made accessible to many. Likewise, if information in not visible, it must not exist, not have been shared, or not be accessible.\footnote{We take the pragmatic view here that if information exists, and is shared and accessible \textit{enough} (contextually defined), it suffices to call that information ``fully visible'' for most work contexts. Our definition of visibility is also distinct from verifiability (especially of authorship) which instead focuses on the ability to the author of a specific section of code \cite{Kalgutkar2019Code}.}

In labor contexts, information availability means that data about the tasks being performed has been created and stored somewhere (e.g. via recordkeeping). Then, those records need to be shared, whether compelled by the law, social norms, or personal conviction, with someone other than the creator. Finally, labor information becomes accessible when an individual knows both where to find the shared information and has sufficient skill to understand what tasks have been performed. Information about labor must meet all three of these conditions --- it must be \textit{\textbf{recorded}}, \textit{\textbf{shared}}, and \textit{\textbf{accessible}} --- in order for it to be \textit{fully} visible \cite{Stohl2016Managing}.

More common perhaps is \textit{partial} visibility, wherein tasks are less documented, less shared, less reachable, or less understandable. Alternatively, tasks may be documented but information only shared, reachable, or understandable to certain individuals (e.g. to experienced group members, but not newcomers) \cite{Fang2009Understanding}. Partial visibility is rarely due to lack of competence; to the contrary, specialized work is often the least visible because experts insulate others from needing to know the steps involved \cite{Daniels1987Invisible, Suchman1995Making}.
% For example, ``writing a paper'' involves many more tasks than putting words on a page. A typical research paper requires knowledge of other researchers' work, identifying a research question or gap, designing a research methodology, collecting data, analyzing the data, knowledge of the norms of publication in a discipline, designing how to present results to that particular audience, etc. Only \textit{then} can we ``write the paper.''

Outside of OSS, academic labor taxonomies like CRediT \cite{Allen2014Publishing, Brand2015Authorship} exhaustively list individuals' activities because, even if a reader witnessed every action that researchers took, readers would have limited visibility into what it took to conduct a study. Likewise, OSS labor records (e.g. the GitHub event stream, octohatrack, and All Contributors) attribute artifact activities, education, management, maintenance, security, and other roles \cite{Young2021Which}. Unfortunately, these systems rarely provide consistent, standardized recordkeeping \cite{Young2021Which} or accessible sharing of those records \cite{Monroy-Hernandez2011Computers}. This creates a need for new measurement techniques as it leaves much labor invisible.

\subsection{Labor Invisibility}
\label{subsec:background_laborInvisibility}

Invisible labor has become a popular term for describing marginalized forms of work both inside and outside the home, like childcare \cite{Daniels1987Invisible}, data entry \cite{DIgnazio2020Data}, and crowd work \cite{Toxtli2021Quantifying}. In one sense, labor becomes invisible if information about the work being done does not exist, is not shared, cannot be reached, or cannot be understood. Any of these conditions is sufficient to make information about labor, and therefore the doing of labor, invisible \cite{Stohl2016Managing}.

Also implied by the concept is that certain tasks tend to be uncompensated or undercompensated \cite{Acker1973Women, Daniels1987Invisible, Babcock2017Gender, DIgnazio2020Data, Toxtli2021Quantifying}. Individuals who perform less-visible tasks may struggle to find financial remuneration \cite{DIgnazio2020Data, Toxtli2021Quantifying}, opportunities for promotion \cite{Riehle2007Economic, Babcock2017Gender}, and to gain social status in their communities \cite{Stewart2005Social}. These activities often fall to women, non-binary people, and other marginalized individuals as a product of gendered socialization processes \cite{Daniels1987Invisible, Berdahl2018Work}. This makes invisible labor particularly pernicious to those who seek greater equity in computing \cite{DIgnazio2020Data, Casari2021Open} and makes \textit{\textbf{compensation}} the fourth determinant of labor invisibility.\footnote{While others define invisible labor in the context of paid employment \cite{Poster2016Introduction, Toxtli2021Quantifying}, they also include activities that are unpaid but expected and, more generally, conceive of invisible labor as ``activities that are tied to a job and its rewards'' and from which an ``employer reaps profits'' \cite[p.7-8]{Poster2016Introduction}. In OSS, formal employers sometimes do and sometimes do not exist, but many parties benefit from OSS regardless (e.g. companies, governments, scientists). For that reason, we argue volunteer activities in OSS ecosystems still merit inclusion, thereby broadening our conceptualization of ``pay'' into ``compensation.''} Together, labor is invisible whenever it is not fully visible, not appropriately compensated, or both (Fig. \ref{fig:conceptual_overview}a).

CSCW has long taken a dialectical view on labor visibility, seeking both to increase visibility while acknowledging the benefits of invisibility. Star \& Strauss \cite{Strauss1985Work, Star1991Sociology, Star1999Layers} coined articulation work\footnote{\textit{Articulation work} is ``work that gets things back `on track' in the face of the unexpected [and] is invisible to rationalized models of work,'' a classic example of support acts \cite{Star1991Sociology}.} in part to raise the visibility of efforts to support distributed work through mechanism design (e.g. new organizing forms, technologies, procedures) \cite{Schmidt1992Taking, Schmidt1996Coordination, Star1999Layers}. At the same time, they agree with Suchman's \cite{Suchman1995Making} arguments that invisibility can also benefit worker autonomy, flexibility, and privacy.

Star \& Strauss' \cite{Star1999Layers} seminal work discusses two relevant concepts for our discussion. First is the concept of disembedded background work. This describes when workers are visible but the work is not, like how a content moderator may themselves be visible, but the emotion regulation they perform while moderating difficult subject matter may not be. With respect to our definition of visibility, disembedded background work implies that labor data does not exist, is not shared, or is not accessible. Star \& Strauss also discuss abstracted and manipulated indicators as another source of invisibility. Here, indicators (like metrics) abstractly describe something that is difficult to measure, meaning the design of these indicators can ``hide'' work and worker by metaphorically distancing decision-makers from work and worker. For instance, having global pools of users, contributors, and maintainers are often viewed as metrics of success for software projects. Yet, numbers of users or countries (designed measures of success) hide the effort required of many individuals to coordinate global open source organizations as they scale \cite{Geiger2021Labor}. This category (abstracted and manipulated indicators) best matches our concept of partial visibility because labor data --- while recorded, shared, and accessible --- does not record, share, or make the practical entirety of an individual's actions accessible.

More recently, Cherry \cite{Cherry2016Virtual}, Ludwig \& colleagues \cite{Ludwig2018Designing}, Gray \& Suri \cite{Gray2019Ghost}, and D'Ignazio \& Klein \cite{DIgnazio2020Data} revealed other instances of invisibility in computing. Content moderation, proof reading, e-waste processing, and countless digitally-tethered activities struggle against the delegitimizing effects of invisibility, suggesting that the detriments of invisibility may outweigh the benefits in many contexts. In effort to measure invisible labor, Toxtli \& colleagues \cite{Toxtli2021Quantifying} recently developed a browser plugin to measure invisible labor in crowd work, finding that Amazon Mechanical Turkers spend nearly a third of their time on invisible tasks. The pervasive effects of invisibility throughout CSCW raises the question of how invisibility manifests in OSS contexts, as well.

\subsection{Invisible Labor in Open Source Software Development}
\label{subsec:background_invisibleLaborOSS}

To understand invisible labor in OSS more specifically, we first need to understand a likely source of tension: the varied motivations of those who participate. We can then turn to evidence of invisible labor in OSS ecosystems and factors shaping its prevalence.

\subsubsection{Motivations}
\label{subsubsec:backgroud_invisibleLaborOSS_motivations}

The most comprehensive review of OSS motivations to date by von Krogh \& colleagues \cite{VonKrogh2012Carrots} sifts OSS motivations into extrinsic, intrinsic, and internalized extrinsic drivers of participation. Extrinsic motivations are those performed to attain external goods, like pay or career opportunities. Intrinsic motivations --- actions performed for their inherent value --- encompass ideological, altruistic, kinship (including community-identification), and enjoyment rationales. Participating for reputation, reciprocity, learning, and own-use value are extrinsic motives that participants internalize and perceive as ``self-regulating behavior rather than external impositions'' \cite[][p.653]{VonKrogh2012Carrots}.\footnote{For detailed definitions of these motivations, see \textit{Carrots and Rainbows: Motivation and social practice in open source software development} \cite{VonKrogh2012Carrots}. We briefly reiterate relevant definitions here for the intrinsic motivations of ideology (that software and its associated practices should be ``free'' and ``open'' to all), kinship (supporting those in a group to which one belongs, like an OSS community), and altruism (selfless concern for the welfare of others, regardless of their group affinity or status). Also relevant are career motivations (achieving objectives of interest to a current or future employer).} Participants often pursue multiple motivations at once, much like OSS communities do \cite{VonKrogh2012Carrots, Sharma2022Motivationhygiene, Smirnova2022Building}. However, motives are commonly in tension with one another because software, social norms, and reward systems differentially advance heterogeneous motivations.

This complicated interplay suggests that individuals might pursue different labor visibility and compensation systems depending on their motivations, which in turn could affect the goal attainment of other participants. For example, one who is motivated by enjoyment alone might not see any benefit in attribution practices, and might even see them as distracting from activities they enjoy. Consequently, they might deprioritize or ignore efforts to increase visibility or compensation. Someone motivated by kinship might see things differently because they seek evidence of their group status (i.e. that they belong or are important to an OSS organization). To them, attribution might confirm their group status, while compensation may or may not. We could similarly conjecture about how visibility and compensation do or do not advance each motivation. Through this lens we can begin to understand evidence of invisibility in OSS.

\subsubsection{Evidence of Invisibility}
\label{subsubsec:backgroud_invisibleLaborOSS_evidence}

Research to date suggests that much of OSS development and maintenance activities remain invisible \cite{Suchman1995Making, Casari2021Open, Geiger2021Labor}. For example, while being an OSS ``maintainer'' typically describes having access permissions to change code, it might also describe how the individual performs strategic decision-making, code testing, version control, financial accounting, or emotionally-intensive activities like user support, community moderation, user request triaging, relationship management, and advertising \cite{Geiger2021Labor}.\footnote{Many of these activities take place ``outside of projects'' --- the traditional activities of creating and modifying code --- but are still embedded within ecosystems.} However, most of these activities remain less visible \cite{Crowston2007Selforganization, DIgnazio2020Data, Trinkenreich2020Hidden, Geiger2021Labor}, receive limited financial compensation \cite{Riehle2014Paid, Trinkenreich2020Hidden}, and involve fluid and often circuitous paths for career advancement \cite{Trinkenreich2020Hidden}, thereby matching the definition of invisible labor.

The invisibility of such activities affects both participants and organizations invested in OSS ecosystems. Participants who seek compensation for work (whether extrinsic forms like pay or career opportunities, or internalized forms like increased reputation) are not likely to receive it without greater visibility \cite{Casari2021Open}. Likewise, organizations cannot understand how OSS ecosystems function (or allocate resources appropriately) without greater visibility into the activities of OSS \cite{Suchman1995Making, Casari2021Open}. Coding examples abound too: projects often combine substantially differentiated coding activities, like bug fixing and feature creation, under the label ``code contributions'' \cite{Casari2021Open}, making granular information opaque behind this catch-all term. So even activities historically valued as ``contributions'' are not fully visible for lack of fidelity, thereby making it difficult to assess the health of projects.

\subsubsection{Factors Shaping Invisibility}
\label{subsubsec:backgroud_invisibleLaborOSS_factors}

Several factors might influence the prevalence of invisibility in OSS ecosystems. First and foremost, conflicting motivations seem a likely source of tension. Individuals may align attribution practices with their own motivations, regardless of how widely held those views are in an ecosystem. Properties of cooperative work technologies might also influence labor invisibility via designed features of software development platforms that either track or ignore certain types of tasks (e.g. articulation work) \cite[c.f.][]{Schmidt1992Taking, Young2021Which}. Individuals might modify their goals through experiences with technology \cite{Gibson2022Sustaining}. For example, if an individual feels they are not receiving fair compensation for their efforts, they might seek to increase the visibility of their work in hopes that greater visibility will yield greater compensation. Finally, community (or ecosystem) governance structures could affect labor invisibility. Governance structures strongly incentivize volunteer participation choices \cite{Shah2006Motivation} with potential to encourage or discourage recordkeeping, sharing, designing for accessibility, and compensating. This may be particularly apparent when stakeholder groups shape norms around conflicting motives, as seen previously in the tension between democratic and bureaucratic governance norms in the Debian ecosystem \cite{OMahony2007Emergence}.

While the literature makes a compelling case for the prevalence and potential causes of invisible labor, funders and researchers still argue for more nuanced, granular, and consistent knowledge of labor activities \cite{Casari2021Open, Geiger2021Labor, Frluckaj2022Gender}. The next section describes how we pursue these objectives through a participatory feminist process and explanatory mixed-methods.

%%%%%%%%%%%%%%%%%%%%%%%%%%%%%%%%%%%%%%%%%%%%%%%%%%%%%%%%%%%%%%%%%%%%%%%%%%%%%%%%%%%%%%%%%%%%%%%%%%%
\section{Methods}
\label{sec:methods}

In this work, we applied a participatory feminist approach \cite{Bardzell2010Feminist, Bardzell2011Feminist} to an explanatory mixed-methods design \cite{Creswell2017Designing}. Our feminist approach identified an articulated need with members of several communities (practicing developer advocates\footnote{The responsibilities of a developer advocate vary, but typically involve interfacing between an organization (which pays the individual) and users of the organization's product. They do this by identifying user needs, advocating to the organization on behalf of those users, and advertising the organization's products to current and potential users.} and OSS community organizers); iteratively co-designed an approach for raising awareness of marginalized experiences in OSS development (both this study and a parallel series of workshops); and shared and discussed our results with members of these communities. Grounded in this approach, the explanatory sequential mixed-methods design \cite{Creswell2017Designing} followed a traditional approach of quantitatively answering a research question (RQ1), posing a new research question to explain the former results (RQ2), and answering the new question with qualitative analysis. The following sections describe our participatory process, how we designed and conducted the survey, and how we analyzed our data.

\subsection{Participatory Process}
\label{subsec:methods_participatory}

To frame our participatory process, we begin with a positionality statement. As coauthors, we identify as English-speaking white academics and practitioners who reside in the West and Global North (United States and Australia). One of us identifies as non-binary, queer, and disabled; one as non-binary; one as a queer woman; and one as a cis man. Our experiences draw from computer science, engineering, and social science disciplines and epistemologies. These intersectional identities and experiences both broaden and constrain how we perceive and construct what it means to ``do work'' and study how people ``do work'' in OSS contexts.

Over about 26 months, we (all coauthors) held a series of biweekly meetings with one another, and intermittently with community members, on the topic of ``OSS Ecosystems.'' During the first 6 months, the topic of these meetings shifted toward ``Ecosystem Mapping'' motivated by a need among our developer advocates to characterize the diversity, quantity, interrelatedness, and vulnerability of tasks that took place throughout several global OSS communities. These meetings led to a series of 8 exploratory digital participatory mapping interviews inspired by physical participatory mapping activities \cite{Chambers2006Participatory, Sletto2009We, Zhou2016Using}. The interviews were designed to understand the breadth, depth, and locales of activities that took place in support of four global Python language communities. In parallel, we conducted two pilot surveys about existing attribution tools that spanned multiple language communities and continents, their efficacies for documenting labor across their respective ecosystems, and developer satisfaction with different attribution techniques. The interviews and surveys led to a series of 6 multi-community workshops, organized by our team, with 33 community members from multiple communities across 5 continents (Africa, Asia, Australia, Europe, \& North America). The workshops supported community members’ efforts ``to identify the work which communities directly attribute to a specific outcome’s success’’ by helping community members define specific outcomes they invest in and identify ``contributions [broadly defined] in a way that is explicit, granular, specific, and transparent.’’

Our preliminary interviews, surveys, and workshops, while not the focus of this article, revealed to us that manual and automated tools have mixed records at measuring diverse activities, including both code and noncode activities, as others before us have shown \cite{Monroy-Hernandez2011Computers, Young2021Which}. This leaves many activities undocumented with repercussions for both individuals and organizations, as described above. Multiple community members expressed a need for better labor ``taxonomies'' and ``metrics'' to demonstrate how much work they do and what that work entails. To us, this resonated with D'Ignazio \& Klein's \cite{DIgnazio2020Data} argument that feminist data work should ``make labor visible'' to benefit marginalized computing workers. Our collective motivation for both the workshops and the design of this survey arose from the desire of community members to make their labor more visible when existing tools did not meet their needs. To that end, we saw need for a critical realist analysis \cite{Terry2017Thematic} (acknowledging a plurality of sociocultural meanings and interpretations of labor and its visibility even as we sought to unearth an underlying ``reality'') to measure the extent of OSS labor, its visibility, and invisibility. We hoped this would yield evidence that individuals of varied backgrounds could provide to traditionally positivist organizations about the extent of their labor.

We therefore co-designed a Qualtrics web survey \cite{Tourangeau2013Science} to assess the pervasiveness of invisible labor. The two academic coauthors on our team drafted questions to measure labor focuses, plus the visibility and compensation dimensions of invisible labor. The full team of coauthors (academics, developer advocates, and community organizers) then iteratively revised all questions and their response options (particularly the artifact and use case categories), both among ourselves and with our Institutional Review Board (IRB), which approved the design before the start of the study. Because we were working with global OSS communities, and given our intent to release the raw data, our IRB recommended making our survey fully anonymous to ensure ethics compliance across multiple jurisdictions. This required us to avoid gathering any demographic data that could identify individuals given the small number of participants in certain demographic cross sections (e.g. an individual of a specific gender, project size, and specific country). The following sections detail the survey instrument's design, distribution, and analysis.

After completing the initial survey analysis, we iteratively prepared a preprint of this article. We shared the preprint with community members to gauge how they felt about our findings, and they in turn shared it with other members of their communities. We plan to circulate this article as a more approachable presentation at developer conferences after peer review.

\subsection{Survey Instrument Design}
\label{subsec:methods_instrumentDesign}

Our critical realist objectives (giving positivist organizations convincing representations of historically marginalized labor) informed our survey design process. To that end, the survey initially focused participant attention on recent collaborative OSS projects (still a more familiar and flexible term for participants than the research concept of an open source ecosystem), before broadening participant perception to the breadth of tasks performed in the ecosystem in which those projects were embedded. This gave us access to participant perceptions of tasks beyond code contributions while still including coding activities. We asked a series of questions about the ``credit'' they receive, which we defined for them as ``the project told someone that you did something for the project,'' as one form of compensation. Then, we asked several questions about how often other people know about the tasks they perform to measure labor visibility. See Appendix \ref{sec:appendix_survey} for the full survey.

\subsubsection{Guiding Participant Attention}
\label{subsubsec:methods_instrumentDesign_participantAttention}

The term ``invisible labor'' and topics related to supportive labor tend to be femininely and liberally coded \cite[c.f.][]{Frizzell2023Timely, Kalita2023How} --- and even denigrated --- in many global contexts \cite{Berdahl2018Work, DIgnazio2020Data}. Socially coded contexts make it difficult to accurately capture lived experiences because participants tend to (a) self-select into or out of surveys and (b) signal through their responses based on how they want to be socially perceived, even if it does not represent their lived experience \cite{Dell2012Yours, Krumpal2013Determinants}. Using coded language like ``invisible labor'' would have increased the likelihood of drawing ideologically aligned participants to the survey; dissuaded ideologically misaligned individuals from taking the survey; and elicited ideologically aligned responses from both groups. Therefore, our goal of accurately representing OSS community members' lived experiences --- regardless of their ideological leaning --- led us to guide participant attention such that we limited gendered and political social signaling. Indeed, \textit{not} limiting social desirability bias would have risked inaccurately representing \textit{all} participants' experiences in aggregate analyses in contradiction with our the critical realist objective of our feminist methodology: ``making labor visible,'' but in a way that positivist organizations also trust.

A participant's experience with a survey begins with recruitment materials. We therefore carefully avoided the term ``invisible labor'' in these materials, along with the consent form (Q1) and other questions throughout the survey. Instead, we identified another term common to open source contexts to help participants recall the full breadth of tasks that people perform, both inside formally-defined projects and outside them (that is, within the ecosystem). We advertised the survey around ``receiving credit'' for their work as this is an important concept to many in open source communities \cite[c.f.][]{Brand2015Authorship, Casari2021Open, Young2021Which} but which should be less likely to evoke either self-selection into the survey based on the survey focus or gendered social desirability bias \cite{Krumpal2013Determinants}. We acknowledge that no terminology is purely neutral nor uncoded, but believed ``credit'' was a common enough term in OSS (even if potentially masculinely coded) that it was less likely to invoke social signaling, insofar as possible while still asking about labor attribution.

The survey began by prompting participants to focus on their experiences with open source projects that they worked on with other people during the last two years (Q2). Initially focusing their attention on collaborative open source projects gave participants the familiar idea of a ``project'' to anchor to (many would not have been familiar with ``ecosystems'' in this context) while giving us access to participant perceptions of OSS ecosystems that pose the greatest challenge to sustainability \cite{NSF2023POSE, NSF2024SafeOSE}. Additionally, focusing their attention on \textit{collaborative} contexts increased the likelihood that participants would more comprehensively consider their actions in the broader ecosystem (the collective organizing context) in which their project (the repository and traditionally related actions) is embedded. We limited the survey to a two-year time period to reduce experience underretrieval due to excessive cognitive load \cite{Lenzner2010Cognitive, Tourangeau2018Survey}.

Later, we focused the participant's attention on how ``tasks vary in shape, size, and how visible they are to other people'' (Q5). We asked them to ``think through the different kinds of tasks you have done as part of these projects over the past 2 years.'' We did this to provide rich retrieval cues about varied tasks and locales, which tend to improve recall \cite{Tourangeau2018Survey}. This broadened their perception out from ``just code'' to include as many tasks as possible within their work, thereby bringing their visible and invisible labor within the broader ecosystem to the fore. For those who already considered non-coding activities to be contributions, this would not affect their judgment; however, for those who might consider code the only legitimate type of ``contribution'' \cite{Casari2021Open}, this design avoided the term contribution (via ``task'' instead) while improving information retrieval about diverse tasks \cite{Tourangeau2018Survey}.

\subsubsection{Compensation Questions}
\label{subsubsec:methods_instrumentDesign_compensationQuestions}

We asked participants about the artifacts and use cases of their OSS collaborations (Q3-4). We did this to measure survey distribution reach, participant diversity, and compensation variation across those settings. After discussion with our IRB, we chose not to include other demographic questions to allow us to distribute the survey globally while protecting varied, nation-specific participant rights through anonymity.

This followed with several questions about receiving credit for tasks (Q6-10). We asked about how many projects they received credit from (via 5-point Likert scale), how many tasks they received credit for (Likert scale), the medium through which they received credit (multiple select), and their satisfaction with the credit they received (Likert scale). These questions helped us gauge how much compensation (here, attribution) individuals receive, how they receive it, and to measure any relationship strength between attribution and perceptions of visibility.

If we want participant responses to approximate frequencies, it becomes important to control for perceptual biases that result from different perceptions of Likert scale terms. Experts suggest linearizing response options (e.g. explicitly numbering response options with 1 through 5 in addition to response text) \cite{Tourangeau2000Psychology} to overcome this. So, we linearized response options for all Likert scales (For example: 1 - Extremely dissatisfied, 2 - Dissatisfied, 3 - Neither satisfied nor dissatisfied, 4 - Satisfied, 5 - Extremely satisfied, I'm not sure). We discuss the visibility questions next.

\subsubsection{Labor Visibility Questions}
\label{subsubsec:methods_instrumentDesign_laborVisibilityQuestions}

Unlike in crowd work where all work might take place on one platform \cite{Toxtli2021Quantifying}, OSS activities take place across an ecosystem \cite{Casari2021Open} that makes measurement via tracking difficult. We therefore measured labor visibility by asking participants how often \textit{different numbers of people} knew about tasks they performed. This approximated how visible different tasks are to OSS communities by asking participants to implicitly assess the extent to which labor data is being created, shared, and made accessible. Combining randomization with human cognitive biases allowed us to estimate likely bounds on how much labor individuals perform at different visibility levels.\footnote{Another alternative could have been to ask participants to identify all the tasks performed (or a representative sample of tasks); about the creation, sharing, and accessibility of data for each task; and to assess sharing and accessibility with respect to each person they interact. However, would induced significant cognitive load as each of these activities requires recalling information on the part of the participant, likely resulting in high drop-out rates. Instead, we opted for the method presented in the text.} We began the labor visibility questions with a prompt (``Think about all the tasks you performed again,'' Q11) to aid recall of their activities. Then, we presented participants with three questions:

\begin{enumerate}[label={Q\arabic*.},font={\bfseries},leftmargin=2cm]
    \setcounter{enumi}{11}
    \item How often did \textbf{2 or more people} know that you performed those tasks?
    \item How often did \textbf{1 other person} know that you performed those tasks?\footnote{We chose to use the phrasing ``1 other person'' instead of several more and less specific alternatives. We did this because in open source contexts some participants may believe that everyone's work \textit{should} be visible and therefore accessible. ``Only 1 other person'' and ``just 1 other person'' may invoke social desirability effects as participants may perceive the words as implying that their work \textit{should} be more visible than it is, thereby shaming them into underreporting. Likewise, the word ``exactly'' provides a rich cue that might encourage participants to second guess themselves. However, using merely ``1 person'' comes at the risk of ambiguity as to whether or not cases in which $\geq2$ people know about a task are subsets of cases in which 1 person knows about a task. We reasoned that ``1 other person'' provided sufficient specificity without risking social desirability bias. Furthermore, we enabled the ``back'' button in the survey so participants could revise their answers if they inferred a different meaning after viewing all three questions ($\geq2$, 1, 0).}
    \item How often did \textbf{nobody else} know that you performed those tasks?
\end{enumerate}

There are a few potential shortcomings to mitigate with this approach. For one, it may be more important for certain individuals to see some information than for others. However, this is often subjective, highly contextual, and difficult to estimate through a self-reported survey. Instead, we remain agnostic to the importance of information to specific individuals by asking about visibility to \textit{any} participants at all with our method. For another, behavioral frequency impressions (e.g. how often someone performs invisible tasks) are prone to under- and overestimation \cite{Conrad1998Strategies}. While we cannot directly verify their perceptions, our approach treats participant perceptions as expertise \cite{Bardzell2010Feminist}. Third, information presented earlier in a survey tends to produce \textit{assimilation effects} --- that is, later judgments are moved in the direction of earlier judgments \cite{Tourangeau2018Survey}, reminiscent of the anchoring and adjustment heuristic or priming effects \cite{Kahneman1982Judgment, Molden2014Understanding}. So if we only presented Q12, Q13, and Q14 in that order (called $\geq2$, 1, 0 hereafter for clarity), participant responses would have been likely to bias results toward presumptions of high visibility (2 or more people, or ``openness''). Therefore, we randomized whether participants saw these questions in order of what we call \textbf{\textit{anchoring to high visibility}} ($\geq2$, 1, 0) or \textbf{\textit{anchoring to low visibility}} (0, 1, $\geq2$).

Following these, we asked participants how important it is to receive credit for tasks (Q16) along with several free response questions about what did and did not work well for them about how people give credit (Q17-19).

\subsection{Survey Procedures}
\label{subsec:methods_procedures}

To test the survey instrument and measure invisible labor in OSS, we recruited 142 participants between January and June, 2022. We did this by posting IRB-approved recruitment materials on the social media platform Twitter with the hashtag \#opensource. Retweets confirm that the recruitment messages reached open source participants on at least 5 continents (North America, South America, Europe, Africa, \& Australia) and a number of OSS developer communities (e.g. Python, Django, pan-African OSS, open science). Per our explanatory design, we answered RQ1 through four quantitative analyses and RQ2 through a subsequent thematic qualitative analysis.

\definecolor{Victoria}{rgb}{0.247,0.282,0.552}
\definecolor{Mischka}{rgb}{0.803,0.811,0.87}
\begin{table}
\centering
\caption{Example subset of participant responses to survey free response questions by participant ID with corresponding codes.}
\label{tab:example_responses}
\begin{tblr}{
  width = \linewidth,
  colspec = {Q[70]Q[700]Q[130]},
  hline{1} = {0.5pt},
  row{1} = {Victoria,fg=white},
  row{2} = {Mischka},
  row{6} = {Mischka},
  row{10} = {Mischka},
  cell{2}{1} = {c=3}{0.943\linewidth},
  cell{6}{1} = {c=3}{0.943\linewidth},
  cell{10}{1} = {c=3}{0.943\linewidth},
}
\textbf{Par. ID}
  & \textbf{Response}
  & \textbf{Codes} \\
Q17 - Think about how the projects you worked on gave people credit. From your perspective, \textbf{what worked well} about how they gave people credit? & & \\
P008
  & \textit{Release shout-outs and thanks are nice.}
  & {announced,\\ genuine thanks} \\
P025
  & \textit{The personal touch. An actual human commenting/noting a contribution is much more personally significant than an automated tool that merely collates automated metrics.}
  & manual, attributed, personal \\
P084
  & \textit{Git commit history for linux kernel is ``permanent'' people can always see your contribution *if they look for it*.}
  & {attributed,\\ internal,\\ documented} \\
Q18 - \textbf{What did not work well} about how they gave people credit? & & \\
P008
  & \textit{They misspelled my name on a thank-you plaque, that wasn't fun.}
  & misattributed \\
P025
  & \textit{When credit mechanisms are arbitrary or completely automated. Completely automated tools don't feel like your efforts are actually being noticed; if the recognition is arbitrary, it's difficult to differentiate between "this maintainer has a vendetta against me" and "this maintainer forgot to recognise me".}
  & {opaque,\\ automated, unattributed} \\
P084
  & {\textit{So many people commit code to the linux kernel that no one bothers to look for your commits unless they need to. Until you are a ``top 10'' (or so) commiter to linux kernel, most people won't know who you are or care. }\\\textit{And getting "credit" for commits "wears off" over time - so unless the commits were in the past year (for larger changes) or couple of months for smaller fixes, it doesn't really matter.}}
  & internal, large, wears off \\
Q19 - Is there anything else that you'd like to share with us? & & \\
P008
  & \textit{(no response)}
  & \\
P025
  & {\textit{My perception is that that projects (at least, the ones that have healthy communities) enthusiastically recognise contributions, and project maintainers are acutely aware of the way that a small recognition can matter to contributors. }\\\textit{However, larger organizations neither recognise nor appreciate the value of those contributions; and often have a very entitled attitude to what they can expect from (often volunteer) maintainers.}}
  & {genuine thanks,\\ personal, large} \\
P084
  & \textit{Credit for committing code is over rated. It's so much easier to write new code than it is to debug someone else's code (and post a fix) or review someone else's code.  People should be getting a lot more credit for reviewing code than committing code. It's part of the reason being a subsystem or driver maintainer in the linux kernel matters way more than committing code - they don't always get credit for this role though.}
  & {certain tasks,\\ inconsistent,\\ attributed}
\end{tblr}
\end{table}

\subsubsection{Quantitative Analyses}
\label{subsubsec:methods_procedures_quantitative}

Our first four analyses utilize descriptive statistics, correlation analysis, and linear regressions. We began by examining how many participants work on different artifacts and use cases to confirm diverse labor participation in our study. Next, we examined how often participants receive compensation (specifically credit) and how visible their work is. We then considered the relationships between variables and calculated causal relationships where possible.

We estimate that, very roughly, between 10 million and 100 million individuals participate in open source projects \cite{Trujillo2022Penumbra}. With our sample of 142 participants, a 20\% proportion percentage (based on five or more Likert scale options per question), and a 90\% confidence interval, this gives a 5.5\% margin of error for our survey.

While our results are not a formal psychometric scale, we calculated reliability estimates on the numeric measures of our survey to assess their internal consistency using coefficient omega total ($\omega_{total}$), now considered the most reliable measure, and Cronbach's alpha ($\alpha$), a more traditional measure, along with our analyses \cite{McNeish2018Thanks, Jebb2021Review}. Reliability estimates of our numeric data suggest high internal consistency ($\omega_{Total}=0.84$, $\alpha=0.78$). For intrinsic consistency, contribution roles, survey data, and code for this project are available at the following:
\begin{quote}
    \url{\repourl}
\end{quote}

\subsubsection{Qualitative Analysis}
\label{subsubsec:methods_procedures_qualitative}

Last, we conducted thematic analysis of qualitative responses to answer our second research question: \textit{What factors affect how commonly labor invisibility occurs in OSS ecosystems?} We described textual responses through qualitative coding, iteratively defining themes that emerged from responses to questions about effective and ineffective credit practices. We followed a standard six step process for ``Big Q'' latent inductive thematic analysis \cite{Braun2012Thematic, Terry2017Thematic}. A ``Big Q'' approach ``operates within a qualitative paradigm and is characterized by (genuine) theoretical independence and flexibility, and organic processes of coding and theme development'' \cite[][p.7]{Terry2017Thematic}. In other words, one researcher subjectively (but iteratively) familiarized themselves with the data, generated codes, constructed themes, revised themes, defined themes, and reported results \cite{Terry2017Thematic}. By ``latent'' coding we mean that our codes sought to capture implicit meanings within the text instead of merely capturing explicit meanings through ``semantic'' coding \cite{Terry2017Thematic}. 

We began by reading the 162 qualitative responses to our free response questions: Q17 (69 responses to ``what worked well?''), Q18 (63 responses to ``what did not work well?''), and Q19 (30 responses to ``anything else you'd like to share?''). Inductive coding initially yielded qualities of importance, and frustration, to participants. Some codes were rooted in participant language (e.g. that credit be ``public,'' ''individual,'' ``genuinely expressed''; and that it not be ``inconsistent,'' ``arbitrary,'' or ``hard to find'') while other codes added slight abstraction (e.g. accessible, distributed, opaque, favoritism). Participant descriptions frequently contrasted with one another (some emphasize the necessity of ``manual'' attribution while others sought greater ``automation,'' for example). A second round of coding refined these initial codes, moving toward multidimensional descriptions of recurring concepts such as attribution (e.g. attributed, unattributed, misattributed) and motivation (e.g. career advancement, own-use, altruism, reputation). Table \ref{tab:example_responses} shows examples of these 38 refined codes alongside corresponding participant responses. Recurring concepts gradually became our initial set of 6 themes (specifically: attribution/``giving credit,'' in/accessibility, compensation, motivation, project characteristics, and process). Borrowing the techniques of asking questions and making comparisons from grounded theory \cite{Corbin2008Basics}, though, analytic memo writing led to a higher level of themes that both distinguish and relate these initial themes. We define and report these final themes toward the end of the Results section in answer to RQ2. But first, we turn to the results of our quantitative analyses in answer to RQ1.

%%%%%%%%%%%%%%%%%%%%%%%%%%%%%%%%%%%%%%%%%%%%%%%%%%%%%%%%%%%%%%%%%%%%%%%%%%%%%%%%%%%%%%%%%%%%%%%%%%%
\section{Quantitative Evidence of Invisible Labor}
\label{sec:quantresults}

% We performed four analyses with the survey data. We began by examining how many participants work on different artifacts and use cases to confirm diverse labor participation in our study. Then, we examined how often participants receive compensation (specifically credit) and how visible their work is. Last, we considered the relationships between variables and calculated causal relationships where possible.

Our quantitative analyses show that participants receive credit for roughly half of their work across diverse artifacts and use cases. Participants reported feeling only slightly satisfied with how often they receive credit and the mediums through which they receive it. We then found that visibility anchoring shapes perceptions of the visibility of a person's work. The following sections detail these results.

\subsection{Sample Captures Diverse OSS Labor Activities}
\label{subsec:quantresults_sampleDiversity}

\begin{figure}[b]
  \centering
  \includegraphics[width=\linewidth]{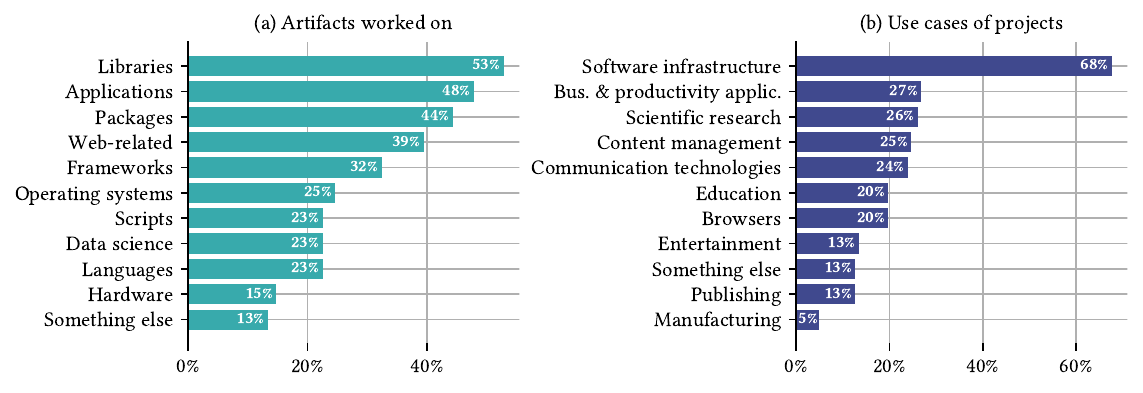}
  \caption{Distributions of (a) artifacts and (b) use cases that survey participants contribute to. Questions were multiple select, so percents shown represent the fraction of participants who expressed that they work on the artifacts and use cases shown.}
  \Description{Horizontal bar graphs displaying distributions of the artifacts and use cases that participants contribute to. At least 30\% of participants expressed that they work on each of libraries, applications, packages, web-related tools, and frameworks. A significant majority (68\%) of respondents work on software infrastructure, and all other use cases are below 30\%.}
  \label{fig:labor_categories}
\end{figure}

Study participants reported working on a wide variety of software artifacts and use cases (Fig. \ref{fig:labor_categories}). Our survey questions on artifacts and use cases (Q3 \& Q4, respectively) each allowed participants to select multiple answers, so percents shown represent the proportion of participants involved in the respective categories.

Each of five categories were selected by more than 30\% of the participants: software libraries, applications, packages, web-related tools, and frameworks. At least 20\% of our sample reported contributing to each of four others including operating systems, scripts, data science, and programming languages. Hardware (15\%) and something else (13\%) saw lower figures, but at least 18 people selected each. Many of the ``something else'' responses consisted of documentation, community building, and research tools.

Project use cases were more skewed among participants. Two thirds (68\%) reported working on software infrastructure, though most selected multiple responses (mean $=2.52$ responses, st. dev. $=1.87$). Six other response options received at least 20\% including business \& productivity applications, scientific research, content management, communication technologies, education, and browsers. Entertainment, publishing, manufacturing, and other uses cases all received less than 15\%. Here, ``something else'' included developer tools, data science, and project management among others.

While our sample of 142 responses clearly cannot represent \textit{all} OSS activity, this sample diversity suggests that trends in our data may reveal broader OSS labor trends.

\subsection{Half of OSS Labor May Not Receive Attribution}
\label{subsec:quantresults_laborCompensation}

\begin{figure}
  \centering
  \includegraphics[width=\linewidth]{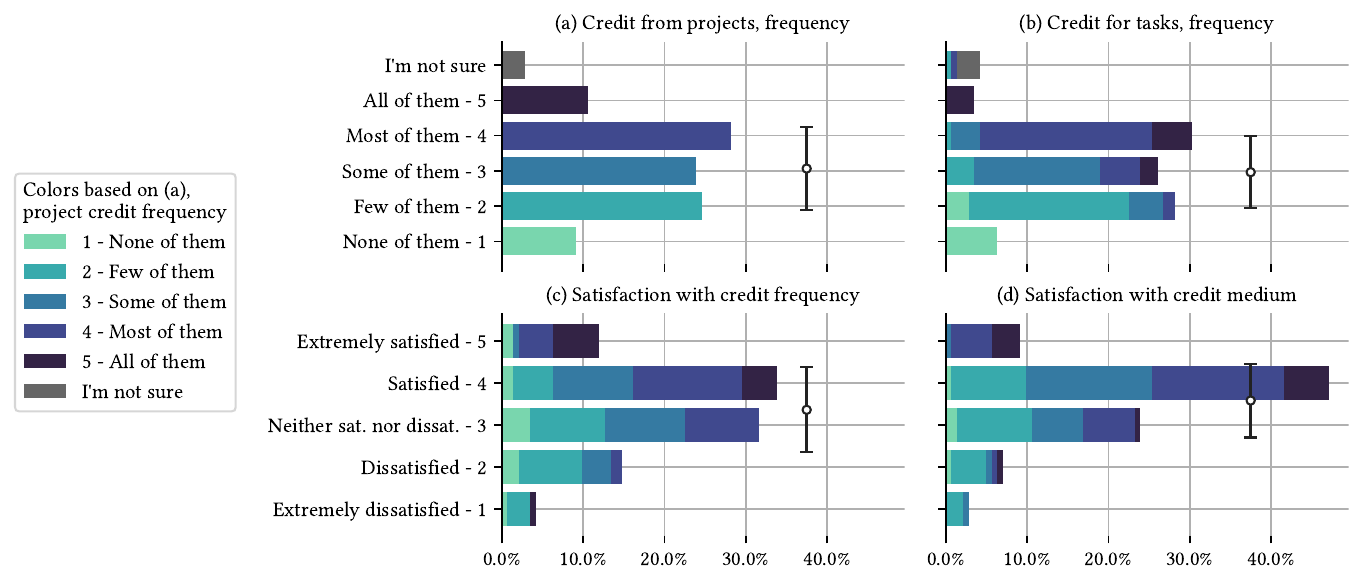}
  \caption{Distributions of responses to questions about how often people receive credit (a) from specific projects and (b) for specific tasks; and distributions for how satisfied individuals are with (c) how often they receive credit and (d) the mediums through which they receive credit. Dots and error bars represent mean values with one standard deviation on the linearized scales. Colors show how individuals responded to (a), tracking those same individuals across the other questions.}
  \Description{The figure shows four horizontal bar graphs. (a) and (b) show wide distributions centered around ``Some of them - 3''; (c) a slightly skewed distribution with most responses on ``Satisfied - 4'' and ``Neither satisfied nor dissatisfied - 3''; and (d) a skewed distribution with most responses on ``Satisfied - 4''.} 
  \label{fig:compensation}
\end{figure}

Next, we examine the data from our compensation questions which focused on receipt of credit for work. The data suggest that participants receive credit inconsistently for the work they do (Fig. \ref{fig:compensation}a-b) but that they still feel moderately satisfied, regardless of how much credit they receive (Fig. \ref{fig:compensation}c-d).

Participants reported that they receive credit from approximately half of the projects they participate in (mean $=3.07$ on a 1 to 5 scale) and for half of the tasks they perform (mean $=2.96$, also a 1 to 5 scale). Significant variation existed among the responses though (st. dev. $=1.17$ for projects, st. dev. $=1.02$ for tasks). Ten percent or less reported receiving credit from projects and tasks either always or never.

In terms of frequency (Fig. \ref{fig:compensation}c) and medium (Fig. \ref{fig:compensation}d), participants reported feeling slightly satisfied with the credit they receive. Satisfaction with credit frequency skewed slightly left, falling just above ``Neither satisfied nor dissatisfied'' (mean $=3.36$, st. dev. $=1.02$). Visual inspection confirms that only about 1 in 5 respondents (19.0\%) felt dissatisfied with how often they receive credit, while about 1 in 3 felt either lukewarm (31.6\%) or satisfied (33.8\%). Then, satisfaction with the medium through which individuals received credit skewed significantly left, falling closer to ``Satisfied'' than ``Neither satisfied nor dissatisfied'' (mean $=3.58$, st. dev. $=0.88$). Here, almost 1 in 2 (47.2\%) reported feeling ``Satisfied'' while 1 in 4 (23.9\%) answered ``Neither satisfied nor dissatisfied.'' Most of this credit came in the form of automated metrics (63.1\%), project documentation (61.5\%), social media posts (50.8\%), and project memberships (47.7\%).

Overall, nearly 9 in 10 (89.1\%) participants reported that some of their work is not fully visible (and therefore invisible) while only 1 in 10 (10.9\%) reported full visibility (Fig. \ref{fig:conceptual_overview}b). While our finding on credit reception frequency cannot speak to other forms of compensation (e.g. pay, social status, opportunities), credit is one of the least financially and socially expensive forms of compensation suggesting that this compensation rate is potentially an upper bound on compensation frequency. Participants' slight satisfaction with credit frequency and mediums may also suggest a possible explanation for why many individuals do not seek more credit. Many who were less satisfied with how much credit they receive listed sources of credit similar to those who were satisfied. However, satisfied individuals cited several sources that dissatisfied individuals cited less often including blogs (47.9\% for satisfied, 5.3\% for dissatisfied), presentations (37.5\% vs. 10.5\%), and project documentation (64.6\% vs. 42.1\%). It is possible, then, that these mediums might increase the efficacy of sharing information or the accessibility of information about OSS labor.

\subsection{Anchoring Shifts Perceptions of Labor Visibility}
\label{subsec:quantresults_laborVisibility}

\begin{figure}[b]
  \centering
  \includegraphics[width=\linewidth]{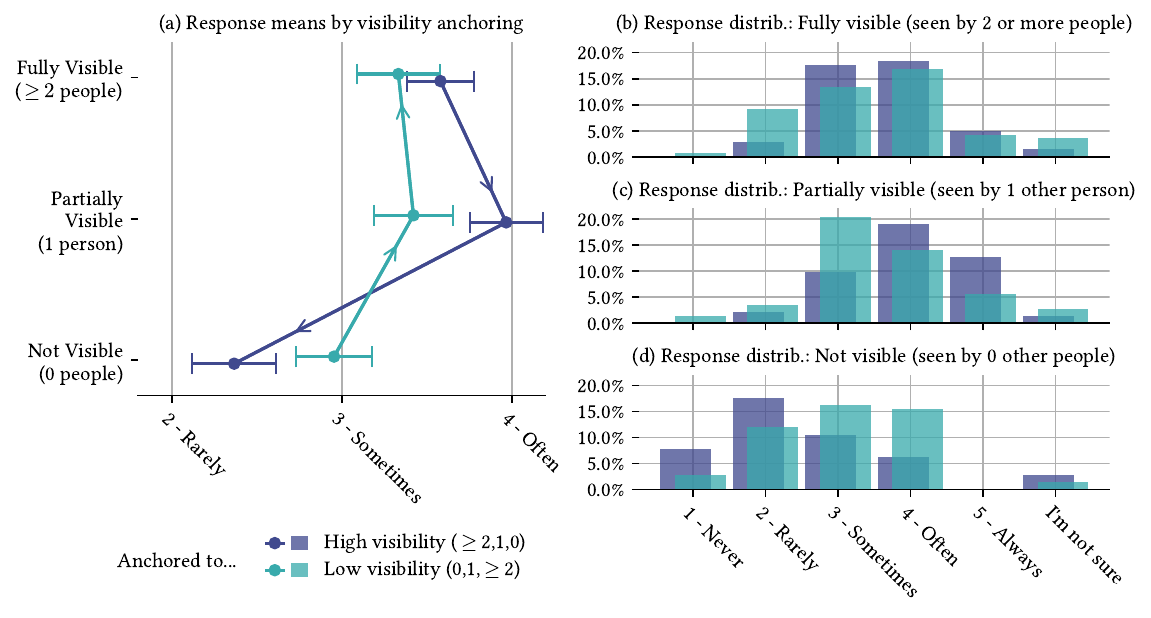}
  \caption{Responses to ``How often did \{2 or more people, 1 other person, nobody else\} know that you performed those tasks?'' as described by response (a) means and (b-d) distributions. Participants who saw the questions in ascending order (0,1,$\geq2$) tended to report that their work is more likely to be invisible (seen by nobody else) and less likely to be partially invisible (seen by 1 other person) than those who saw questions in the descending order ($\geq2$,1,0). In (a) arrows indicate question viewing order, and errorbars show 95\% confidence interval of difference between means.}
  \Description{(a) shows a plot of points with errorbars and connecting lines indicate the question viewing direction. (b-d) show vertical bar graphs of the distributions of responses to each of the two viewing directions.}
  \label{fig:visibility}
\end{figure}

Turning to labor visibility, participants either viewed questions starting from an anchor to high visibility (seen by $\geq2$, 1, then 0 people) or low visibility (0, 1, $\geq2$ people).  Averaging the results of the two groups and placing them in proportion to one another (Fig. \ref{fig:conceptual_overview}b) approximates how often labor is invisible (2 in 3, 66.7\%) compared to fully visible (1 in 3, 33.3\%). But separately, Fig. \ref{fig:visibility} shows that the anchors produced statistically significant differences for the two groups, both in their mean responses on each question (Fig. \ref{fig:visibility}a) and the distributions of their responses (Fig. \ref{fig:visibility}b-d). Together, their answers suggest ranges in which the ``true'' likelihood of labor visibility may fall.

The group anchored to high visibility (shown in purple) viewed questions in descending order. They usually reported that $\geq2$ people knew about their work ``Sometimes'' or ``Often'' over the past two years (Fig. \ref{fig:visibility}b); then that 1 other person would ``Often'' or ``Always'' know about their work (Fig. \ref{fig:visibility}c); and finally that ``Rarely'' or ``Sometimes'' would nobody else know about their work (Fig. \ref{fig:visibility}d). Again, survey respondents tend to answer later survey questions by making adjustments from earlier questions \cite{Tourangeau2018Survey}. This progression suggests the group's relative perceptions of the labor visibilities; to them, their work was most likely to be partially visible (1 person), less likely to be fully visible ($\geq 2$ people), and less likely still to be not visible (0 people). Cumulatively, the cases that satisfy the conditions for invisibility (when labor is either not visible or partially visible) are more likely than full visibility.

The group anchored to low visibility (turquoise) significantly contrasted the previous group. Viewing questions in ascending order, they were more likely to report their work was ``Often,'' ``Sometimes,'' or ``Rarely'' visible to nobody else (Fig. \ref{fig:visibility}d); ``Sometimes'' or ``Often'' visible to 1 other person (Fig. \ref{fig:visibility}c); and ``Sometimes'' or ``Rarely'' again for $\geq2$ people (Fig. \ref{fig:visibility}b). Here too, this group likely made adjustments from their anchor (nonvisibility). Consequently, this group's averages suggest that they perceive all three cases --- full, partial, and nonvisibility --- as happening roughly as often as one another, though partial visibility may occur slightly more often than nonvisibility based on their distinct 95\% confidence intervals (Fig. \ref{fig:visibility}a). Here, too, the cumulative combination of partially visible and nonvisible labor makes invisibility more likely than fully visible labor.

The two groups roughly agree on how often their OSS labor is fully visible, revealing that about 2 in 3 tasks are invisible  (Fig. \ref{fig:conceptual_overview}b). Still, the differences between groups on nonvisibility and partial visibility are significant, statistically and in implication. First, the group anchored to high visibility perceived invisible labor as less common than the low visibility anchored group, suggesting that anchoring may influence perceptions of labor visibility. This appears again in the difference between the partial visibility means: both groups perceived partial visibility as more likely than their respective anchors, but the groups differed on the likelihood depending on the initial framing. The qualitative differences between the response distributions in Figs. \ref{fig:visibility}b \& c further support this.

\subsection{Visibility Anchoring May Shift Beliefs About the Importance of Credit}
\label{subsec:quantresults_anchoringCausality}

\begin{figure}
  \centering
  \includegraphics[width=\linewidth]{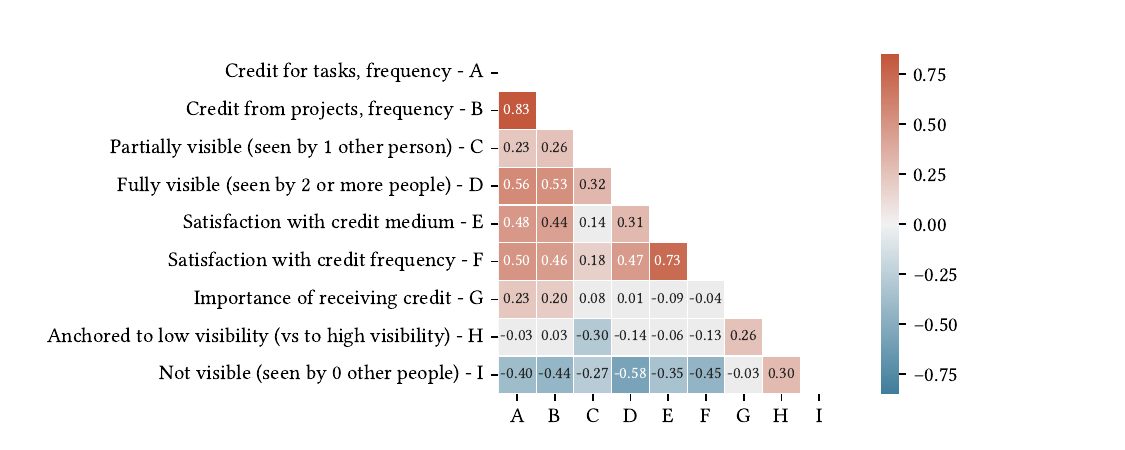}
  \caption{Correlations between numerical survey variables, ordered by clustering. Colored boxes are statistically significant ($p<0.01$) while gray boxes are not. Being seen by nobody else (I) was correlated with low scores on several other variables. Conversely, being seen by $\geq2$ people (D) was often correlated with high scores on other variables. Being seen by 1 other person yielded mixed results.}
  \Description{A heatmap grid of correlation values with variables listed across the vertical and horizontal axes. More positively correlated relationships are colored increasingly darker shades of red; more negatively correlated relationships are colored increasingly darker shades of blue; uncorrelated relationships are colored white; and statistically insignificant relationships are light gray.}
  \label{fig:correlation_matrix}
\end{figure}

Next, we examine relationships between the quantitative variables of our survey. Calculating Pearson correlation coefficients between these variables reveals several trends (Fig. \ref{fig:correlation_matrix}). Unsurprisingly, the likelihoods of receiving credit for specific tasks and from specific projects are highly positively correlated ($r=0.87$), as were participants' satisfactions with the credit they receive ($r=0.73$, compare Fig. \ref{fig:compensation}c \& d). Credit frequency and participant satisfaction tended to correlate with one another (A, B, E, \& F) and with full visibility (D). Partial visibility (C) also correlates with the former but less so, while negatively correlating with anchoring order (H, $r=-0.30$) and being not visible (I, $r=-0.27$). Indeed, negative correlations proved common for nonvisibility (A-F) and only positively correlated with anchoring to low visibility (H, $r=0.30)$. Anchoring to low visibility also correlated with importance of receiving credit (G, $r=0.26$), which in turn correlated with both credit frequencies ($r=0.23$ \& $r=0.20$, respectively).

\begin{figure}
  \centering
  \includegraphics[width=\linewidth]{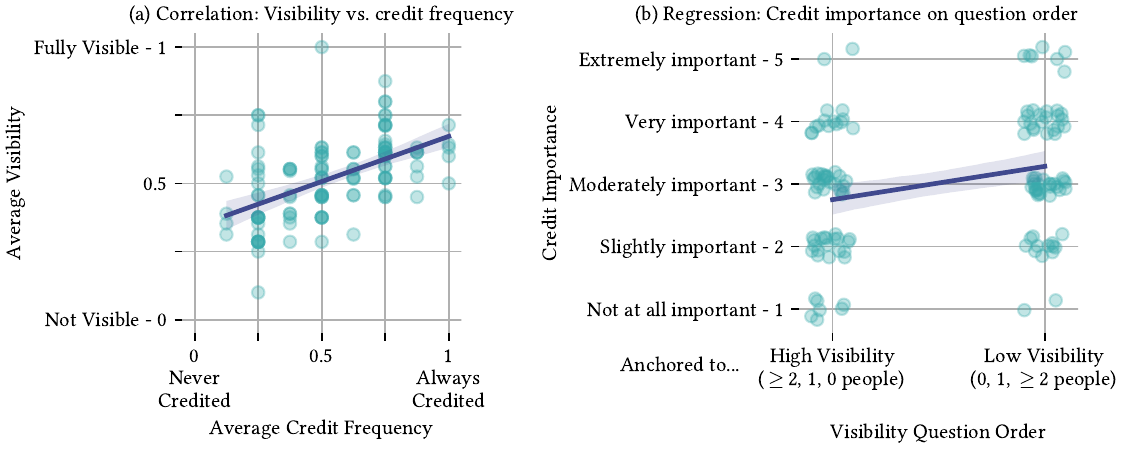}
  \caption{(a) Scatterplot of average visibility versus average credit frequency. Points indicate individual responses. The regression line shows the correlation between variables. (b) Linear regression of credit importance against visibility anchoring. Points indicate individual responses and are grouped on the x-axis by presentation order. The regression line of best fit shows a statistically significant positive relationship between labor invisibility (via anchoring) and greater importance of credit (Adj. $R^2=0.061$).}
  \Description{Two scatterplots, one of average labor visibility versus average credit frequency with a positive trend. The second regresses how important receiving credit is to participants against question presentation order. Here, the x-axis is broken out by presentation order (decreasing on the left, increasing on the right) and the y-axis shows importance responses from 1 (not important) to 5 (extremely important). A regression line of best fit is overlaid showing a slight slope upward from left to right.}
  \label{fig:variable_relationships}
\end{figure}

Again, invisible labor consists of both visibility and compensation. To understand the relationship between these elements in OSS (here, with credit as compensation), we calculated a weighted average of the visibility measures (fully, partially, \& nonvisible) and separately averaged the frequencies of receiving credit (credit for tasks, credit from projects). Fig. \ref{fig:variable_relationships}a shows the relationship between average visibility and average credit frequency for each participant ($r=0.53$). While this is correlative and not causative, visibility and credit clearly trend together. Few individuals' responses significantly deviated from this trend. Calculating the average of averages for each dimension provides further confirmation that, in general, OSS labor is only partially visible (51.9\%) and roughly half receives credit (50.1\%).

The trend between visibility and credit becomes more important in the context of one last observation. Recall that the importance of receiving credit (Fig. \ref{fig:correlation_matrix}, G) correlated with anchoring order (H). The importance of receiving credit (Q16) appeared \textit{after} the visibility questions (Q12-14) meaning that anchoring order had the potential to yield \textit{causal} effects for subsequent questions. Fig. \ref{fig:variable_relationships}b shows that anchoring had a statistically significant effect on how important credit was to participants. On average, participants anchored to low visibility thought receiving credit was $30.5\%$ more important than participants anchored to high visibility ($95\%$ confidence interval of $\pm19.5\%$, $p<0.002$, Adj. $R^2=0.061$ with robust standard errors).

To summarize, we found that labor is only partially visible and compensated on average (each about 50\%). Further, we found that visibility anchoring shapes perceptions of the visibility of a person's work. Asking participants about highly visible work (seen by $\geq2$ people) first in our series of questions increased participant frequency perceptions of partially visible work (seen by 1 other person) and decreased frequency perceptions of low visibility work (seen by nobody else) compared to those whom we asked about low visibility first.

%%%%%%%%%%%%%%%%%%%%%%%%%%%%%%%%%%%%%%%%%%%%%%%%%%%%%%%%%%%%%%%%%%%%%%%%%%%%%%%%%%%%%%%%%%%%%%%%%%%
\section{Qualitative Explanation of the Commonness of Invisible Labor}
\label{sec:qualresults}

Given this quantitative evidence, we now consider factors that affect how commonly labor invisibility occurs (RQ2). Our free response questions asked participants about their experiences with how projects gave credit (Q17-19). In what follows, we analyze responses to these questions to identify factors that affect the prevalence of invisible labor, treating expressed experiences as expertise \cite{Bardzell2010Feminist}. We first corroborate our quantitative findings, providing qualitative evidence of labor invisibility from our sample. Then, our thematic analysis identifies a theme of \textbf{\textit{attributing at cross-purposes}} --- pursuing diverse motivations through conflicting attribution practices with limited resources --- and its influence on the pervasiveness of invisible labor.

%%% Evidence of Invisibility %%%

\subsection{Experiences of Invisibility in OSS Ecosystems}
\label{subsec:qualresults_invisibility}

Generally, participants written responses expressed dissatisfaction with the extents of visibility and compensation in OSS ecosystems, the constituents of invisible labor. A few respondents dissented from this view, though, expressing either apathy or hostility toward the idea of receiving credit. We address each respectively.

\subsubsection{Visibility}
\label{subsec:qualresults_invisibility_visibility}

Participants often recounted the non- and partial visibility of OSS labor. For example, participant P110 described how ``contributions are seen by a handful of people at most which makes contributions almost invisible.'' Similarly, P002 described how the work they do ``only shows in project-specific mediums via being parsed out of commit messages, [which] doesn't help with other things people might care about such as GitHub graphs.''

This second response exemplifies how technologies both enable and constrain visibility. Here, P002 implies that ``GitHub graphs'' achieve better visibility than ``commit messages.'' Interpreted through the elements of visibility (being \textit{recorded}, \textit{shared}, and \textit{accessible}; see Sec. \ref{subsec:background_laborVisibility}), P002 perceives a labor visualization tool (the GitHub graph) as increasing labor data accessibility for external stakeholders in a way that parsing one's name from commit messages does not. This matches our understanding that recordkeeping alone, without sharing and accessibility of labor information, does not yield full visibility. That said, other participants took issue with the counting function performed by visibility tools like the GitHub graph:
\begin{quote}
    \textit{Counting tasks or git issues can be super misleading. People can comment on an Issue and get credit just for participating, which should be appreciated. But there is no good solution for people who spend several hours writing an Issue or an Issue comment. Time spent on a task is not considered, nor are skill levels. Often programming/coding is seen as the only valid contribution type, and everything else is seen as trimmings and less valuable.} (P104)
\end{quote}
Another participant echoed these sentiments: ``github showcases contributors to a repository in the sidebar, but only commits count as contributions, whereas activities like issue reporting, issue triaging, code review, translations performed in third-party platforms, spreading the word, etc., tend to be overlooked'' (P82).\footnote{GitHub has gradually expanded the activities that count toward the contribution graph to include issues, pull requests, discussions, and some commits. But we do not have evidence of the exact dates of when these changes were introduced, and to date it cannot account for activity outside the platform.}

These quotes demonstrate how some OSS labor gets ``overlooked'' (i.e. recorded and shared less often), consistent with the literature \cite{Geiger2021Labor}. Activities such as ``issue reporting, issue triaging, code review, translations \ldots spreading the word, etc.'' are only ``seen by a handful of people at most.'' This happens partially because certain activities are ``seen as trimmings'' (i.e. not as \textit{real} work) and as ``less valuable.'' Additionally, technologies make certain ``valid contribution[s]'' (especially ``programming/coding'') more accessible by ``showcas[ing]'' them, although technologies also misrepresent certain dimensions of even legitimated activities (time, skill). Collectively, these choices have the effect of ``misleading'' people (both internal and external) about labor that takes place in an ecosystem. Put another way, decisions about what we make visible via recordkeeping, sharing, and accessibility often misrepresent the full extent of what gets done.

\subsubsection{Compensation}
\label{subsec:qualresults_invisibility_compensation}

Many participants also felt that OSS work received insufficient compensation (mostly referring to credit, but not only). Several participants described not receiving compensation on the basis of an individual's status, such as community tenure status or residency status. On community tenure, for example, P024 responded with frustration that a project ``didn't [give me credit] at all. I made my first contribution to a browser and I was hoping to be listed anywhere, even if buried in some page of the browser's system that no one ever finds organically.'' By implication, they were not listed in those buried pages. On residency status, P009 described an experience common to many in minoritized regions who participate in global ecosystems: ``Some people are less likely to receive credit based on timezone/location. For example, when Americans are giving out special credit, they tend to forget about Asians because they don't see them in meetings a lot. Also, company loyalty messed up crediting people on multiple occasions.''

Others described their compensation experiences in starker terms: ``It's not just a credit/no credit issue, but a credit/no credit/\textbf{\textit{erasure}} issue'' (P131, emphasis added). While ``erasure'' is a visibility metaphor, the juxtaposition of ``erasure'' with ``credit/no credit'' here implies that erasure refers to \textit{taking away} something owed, something compensatory that participants deserved. Depictions of ``erasure'' were common. One participant described how ``maintainers would \textit{take credit} for things they hadn't done, and refrain to give any but bare credits and acknowledgement for other people's work'' (P129, emphasis added). Another described how ``after 18 months of work on a project, nobody from the project leadership gave me any attribution or thanks'' (P027). A third described how ``people take the work and before it has your name on it, do some tangential re-spin, and commit in their name. What annoys me isn't really the credit, but they act like a bad client, who isn't paying you'' (P062). A less generous interpretation might be one of theft. Another participant made this explicit through a metaphor of ``slavery'':
\begin{quote}
    \textit{Credits are good. But at some point you realize that while you're being credited and such, you live with parents, and barely make \$150 a month to compensate them your living costs. Because by contributing to OSS and getting credit you're in a reward loop that makes all proprietary work look like slavery under a mighty boss.} (P101)
\end{quote}

Participants described insufficient compensation in terms of harm. In their words, it ``hurts people when it's inconsistent, and not everyone is recognized in the same way'' (P134). Some harms compare to affective pressures found in many workplaces: ``In some of these scenarios where your role is not just creating a PR [Pull Request] with code contributions, I sometimes felt a pressure to keep up and make sure my role was still useful to the team'' (P099). Other times, these pressures may have devolved into harassment: ``Contributing contentious kind of work (design and Product management) is hard and very hostile at worst'' (P096).

Beyond individual harms, giving insufficient compensation limits governing organizations, too. Labor retention grows from both giving credit (``Giving credit is important. Not giving credit can drive people away quickly,'' P086) and giving it fairly (``Too many projects give credit to a small group [of workers], making it hard for the project to bring in new people,'' P018). Some ecosystem leaders took this responsibility quite seriously, describing how they ``deliberately try to acknowledge the work of others even if it downplays my own efforts. People need thanks. I used to send Christmas cards to the top contributors or new ones that needed a thanks'' (P086). In sum, many participants felt insufficiently compensated and hurt as a result.

\subsubsection{Dissenting Perspectives}
\label{subsubsec:qualresults_invisibility_dissent}

Not all participants cared about labor invisibility, though. Some expressed relative indifference. They mentioned that ``receiving credit for contributions to open source projects is certainly nice and makes me feel good, but it's lower on the list of reasons why I contribute'' (P040); that ``this isn't really something that concerns me..'' (P045); and that ``I do it because I want it, do not need credit'' (P116). Others were more blunt:
\begin{quote}
    \textit{If you come across anyone contributing to open source where recognition is important, they are the problem with open source. I contribute because I want to, because I need the feature, or because I want to see if I can do something. That's how OS started. I stop contributing when I want to for my own reasons, not because I'm not getting enough street cred or recognition.} (P020)
\end{quote}
While a minority perspective, this participant was not alone in dissenting on the basis of personal enjoyment, learning, and own-use motivations.

In short, most participants described portions of their work as invisible, either for being insufficiently visible or insufficiently compensated. Not all care (tying into motivations), but most do. These descriptions provide a backdrop for our thematic analysis, which we turn to next.

%%% Thematic Analysis %%%

\subsection{Attributing at Cross-Purposes}
\label{subsec:qualresults_attribution}

Shaped by our survey questions, our thematic analysis identified one factor in our dataset that appears to influence the prevalence of invisible labor in OSS ecosystems --- attributing at cross-purposes  --- with three subthemes related to participant motivations and attribution practices (Fig. \ref{fig:attribution_triangle}).\footnote{Other factors surely also influence labor invisibility. Here, we report only factors that we had evidence for as viewed through the theoretic lenses of this work. We discuss other possibilities in the Background (Sec. \ref{subsubsec:backgroud_invisibleLaborOSS_factors}) and the Limitations (Sec. \ref{subsec:discussion_limitations}).}

Again, motivations for participating in OSS vary and often coincide with one another (see Sec. \ref{subsubsec:backgroud_invisibleLaborOSS_motivations}). Our participants' motivations were no exception. Affinity for community (i.e. kinship), career advancement, and pay were most prominent in our responses, though we also found evidence of interest in helping others (altruism) and improving tools (own-use). These motivations were often latent within responses about credit qualities that participants liked or disliked. Consider an excerpt from P062: ``I guess it's nice to have visible [data] to show when looking for jobs or in a negative phase looking for things to be positive about.'' This excerpt contains an extrinsic career motive (``when looking for jobs'') tied to accessible labor information (``have visible [data] to show''), and a further unspecified motive (``looking for things to be positive about''). This latter motive is likely either intrinsic (e.g. altruism, kinship, fun) or internalized extrinsic (e.g. learning) because such motives tend toward greater affect \cite{VonKrogh2012Carrots} (as evidenced by the words ``be positive about'').

While not mutually exclusive, individuals' actions in pursuit of these motivations were often at cross-purposes with one another, though. By this, we mean that individuals and organizations might earnestly pursue several motivations simultaneously through attribution designs, but that such pursuits often got ``complicated'' (P004 \& P058, separately) because efforts to advance one motivation frequently came at the expense of effectively pursuing another. This tension was particularly evident between three distinct forms of attribution (our subthemes) that participants spoke of: expressive attribution, instrumental attribution, and non-attribution.\footnote{We draw our terminology for attribution from Balkundi \& Harrison's work on social network tie content \cite{Balkundi2006Ties}, though we acknowledge that other terminologies exist within CSCW to describe related concepts. Specifically with respect to Monroy-Hern\'andez \& colleagues \cite[e.g.][]{Monroy-Hernandez2011Computers}, our instrumental attribution subtheme partially overlaps their ``attribution'' category, and our expressive attribution subtheme closely aligns with their ``credit'' category. We do not see a parallel to the non-attribution subtheme.}

The first type, \textit{\textbf{expressive attribution}}, describes genuine, personalized thanks for actions performed. This form of attribution is fundamentally relational in that individuals seek and appreciate sincere, person-to-person, non-automated expressions of affective states about labor performed. Those who sought it wanted not just ``proper attribution of what was performed, [but also] why it was helpful'' (P102). Expressive attribution likely provides individuals with ``personal validation'' (P045) of successfully fostering kinship, a fundamentally relational motive. At the same time, terms like ``proper attribution'' demonstrate how they did not view the giving of personalized thank you messages as distinct from internalized or extrinsic motivations. Sometimes participants sought public announcement of personalized statements (reputational motivation), for instance, which might yield invitations to be a ``guest on a podcast, co-present at a conference, write a blog post'' (P100). But in general, expressive attribution did not inherently require visibility so much as \textit{genuine thanks}.

Next, \textbf{\textit{instrumental attribution}} strives to systematically associate labor data with individuals who performed specific actions, and to make the data visible to those who might benefit from it. In contrast to the relational objectives of expressive attribution, this type of attribution is bureaucratic --- not in the popular sense of \textit{needless} process, so to speak, but rather because it seeks \textit{effective processes} that systematically and transparently attribute labor, regardless of one's relationship with others. Here, people often wanted attribution to be given ``in an automated or semi-automated fashion'' (P082) such that ``everybody got credited, regardless of task/relevance/work, etc.'' (P044) in a ``specific and accurate'' way (P042). Instrumental attribution seems to support altruism (i.e. ``everybody got credited''), ideology (i.e. that work \textit{should} be open), own-use (organizational control), and extrinsic (career, pay) motivations. It does this by systematizing recordkeeping, information sharing, and labor data accessibility in support of varied participant motivations.

Third is \textbf{\textit{non-attribution}}, a type which was less common in our sample. Nevertheless, a few participants noted that they do not need (or even want) attribution. These individuals described their participation as ``I do it because I want it [the software], do not need credit... I do it because I like it and not because others need to like it'' (P116). These individuals see attribution as superfluous, unnecessary, or at least a lower priority. They self-describe as being motivated by own-use and fun. While labor visibility could hypothetically be an activity individuals associate with these motivations, we did not see evidence of it in our sample.

Participants expressed frustration at not receiving attribution in the way they wanted (i.e. the attribution type that best supports their motivations). Those seeking the genuine thanks of expressive attribution disliked the automated systems of instrumental attribution because they felt disingenuous (``anything automated feels fake,'' P088), insufficiently personalized (``credits just don't help to see how much damn effort it takes,'' P101), and often produce inaccessible labor data (``Change log [attribution] wasn't really useful,'' P079). Those interested in instrumental attribution worried about how non-attribution hid work (``there was no way for others to see what had been done or not,'' P057); for expressive attribution, they disliked the inconsistency (``projects give credit by accident, if at all,'' P069), favoritism (``tends to be a few who garner all of the credit,'' P112), and labor intensiveness (``often it's a manual process that's not easily scalable,'' P082). Advocates of non-attribution expressed frustration with any attention to attribution (``anyone contributing to open source [for whom] recognition is important, they are the problem,'' P020).

To reconcile this discord --- particularly between proponents of expressive and instrumental attribution --- many participants described mixed automated-and-manual attribution designs (``Mentioning in releasnotes [sic] and announcements on the blog,'' P070). The objective therein was to both give personalized thanks to help people feel valued (kinship motivation) \textit{and} thorough, accessible documentation that individuals could point to for career advancement. On the other hand, one person noted that their organization is discussing ``whether individuals WANT to be named or credited --- credit described in documentation offers that flexibility whereas automated systems would not account for this nuance and have the potential to extend beyond the comfort levels of participants'' (P087). These balanced aspirations commonly gave way to pragmatic concerns, though:
\begin{quote}
    \textit{Credit systems that give credit for all types of work, such as design, project management, event organizing, etc. [are ideal, but] too much manual effort required to give credit. Difficult to find mentors through the credit system. Hard to see who is doing what work.} (P117)
\end{quote}
\begin{quote}
    \textit{The best situation is when the project is open to accept different methods of giving credit. For example, when in the past you needed 2 commits (in code) to get funds for a hackfest, now you can get funds even if you're not a programmer and you contribute with marketing skills, community management etc. The most reliable metrics are stored in the memory of people who constantly hand out and know what's happening. If they drop out, so does a lot of project value/knowledge/processes, and past credit they observed of other people.} (P104) 
\end{quote}
From this, we see that limited individual resources (e.g. time, attention, knowledge, funds) and collective resources (like skilled members, accessible labor data, and capable technology) appear to constrain efforts to pursue everyone's motivations to the fullest, here through attribution.

Altogether, this leaves OSS ecosystems in tension:
While some participants \textit{do} want greater visibility and compensation (instrumental attribution), others do not.
Instead, they want to feel like part of a community (affirmed by expressive attribution) or to have fun doing whatever tasks they take pleasure in (non-attribution).
Limited organizing, financial, and technical resources then mean that leaning into one type of attribution may help those who share similar motivations (as advanced by a particular form of attribution), but at the cost of pursuing or facilitating others.
In this sense, efforts supporting one type of attribution (and the corresponding motivations) might not advance, or may even hinder, efforts to support other types of attribution (and their motivations).
True, multiple attribution types can (and do) exist at once, but this too draws on limited resources, subsequently limiting abilities to pursue each form of attribution (and, once again, their motivations).

\begin{figure}[t]
\centering
    \begin{tikzpicture}[scale=0.8]  % Slightly increased scale for better visibility
        \coordinate (A) at (0,0);
        \coordinate (B) at (2,0);
        \coordinate (C) at (1,{-2*sqrt(3)/2});
        
        \draw (A) -- (B) -- (C) -- cycle;
        
        \node[anchor=south east, align=right, text width=2cm] at ($(A)+(-0.1,0.1)$) {Expressive\\ Attribution};
        \node[anchor=south west, align=left, text width=2cm] at ($(B)+(0.1,0.1)$) {Instrumental\\ Attribution};
        \node[anchor=north] at ($(C)+(0,-0.1)$) {Non-Attribution};
        \filldraw ($(A)+(0.282,-0.175)$) circle (2pt) node[anchor=north east] {A};
        \filldraw ($(A)!0.5!(B)$) circle (2pt) node[anchor=south] {B};
    \end{tikzpicture}
    \caption{Attribution types are often at cross-purposes with one another because pursuing motivations through one type of attribution (expressive, instrumental, or non-attribution) often constrains efforts to pursue other motivations. The triangle therefore symbolizes an ``attribution space'' in which one's actions fall. For example, the more effort one puts toward expressive attribution (point A), the more difficult instrumental and non-attribution become. Likewise, equally supporting expressive and instrumental attribution (point B) for everyone would require significant effort and might frustrate those who prefer non-attribution.}
    \Description{The figure shows an equilateral triangle with the points of the triangle labeled ``expressive attribution,'' ``instrumental attribution,'' and ``non-attribution,'' the three subthemes of cross-purposed attribution.} 
    \label{fig:attribution_triangle}
\end{figure}

Therefore, the confluence of diverse motivations and resource limitations frequently puts the attribution practices of individuals and organizations at odds with the motivations of others;
that is, participants find their attribution practices at cross-purposes with one another.
This includes efforts to reduce labor invisibility (primarily through instrumental attribution) which often limit efforts inspired by the motivations of other participants (as embodied in expressive and non-attribution).
We represent this tension as a triangle delimiting the ``attribution space'' of possibilities (Fig. \ref{fig:attribution_triangle}).
This symbolically represents how efforts to pursue motivations through one or more attribution types (moving toward those points on the triangle) draw limited resources away from pursuing motivations through other attribution types.
The triangle metaphor visualizes how the tension between attribution practices decreases labor visibility and compensation because many participants are motivated by interests that do not benefit from, and may even be hindered by, greater attention to labor visibility and compensation.

In this light, our finding for RQ1 that roughly half of OSS labor remains invisible becomes less surprising. When OSS managing organizations design attribution systems (however formally or informally), those systems differentially advance some motivations more than others (answering RQ2). Now we turn to the implications of these findings.

%%%%%%%%%%%%%%%%%%%%%%%%%%%%%%%%%%%%%%%%%%%%%%%%%%%%%%%%%%%%%%%%%%%%%%%%%%%%%%%%%%%%%%%%%%%%%%%%%%%
\section{Discussion}
\label{sec:discussion}

%%% Summary
As we have seen, labor becomes invisible labor when it is undercompensated or not fully visible (Fig. \ref{fig:conceptual_overview}a). Through a labor-diverse sample of OSS ecosystems, we demonstrated that invisible labor makes up about half of the work taking place in OSS ecosystems (Fig. \ref{fig:conceptual_overview}b) and that visibility anchoring shapes perceptions of the visibility of a person's work. Resource limitations seem partially responsible for labor invisibility as evidenced by our expos\'e of the tension between attribution motivations (Fig. \ref{fig:attribution_triangle}). We now discuss implications in the context of those who came before us, along with the work's limitations and future work.

\subsection{Implications}
\label{subsec:discussion_implications}

%%% Implications --- 50% Invisible
At some level, the commonness of invisibility in OSS is not surprising; researchers have recounted this since the inception of CSCW \cite{Suchman1995Making, Star1999Layers}. At the same time, \textit{showing} that such a significant proportion of work is invisible, as we see elsewhere in CSCW \cite{Toxtli2021Quantifying}, holds mixed potentials. Invisibility often shrouds valuable work expertise \cite{Suchman1995Making, Star1991Sociology}. So on one hand, our findings could increase the legitimacy of articulation work by corroborating the abundance of transferable knowledge embedded in OSS ecosystems \cite{Jergensen2011Onion}. On the other hand, that 2 in 3 tasks are not fully visible reminds us that these critical ecosystems \cite{NSF2024SafeOSE}, and less powerful participants in them \cite[c.f.][]{Star1991Sociology, Star1999Layers}, remain vulnerable. We found that work was most likely to be partially visible (seen by 1 other person), so if the singular external viewer of a task loses visibility into that task --- by promotion, leaving the community \cite{Jergensen2011Onion}, or even just forgetting --- that labor becomes unknown to others. Second, some labor is \textit{already} seen by nobody other than the worker as we showed. Either way, these tasks are invisible and therefore difficult for others to understand, compensate, and manage.

The abundance of studies that struggle with OSS labor and its invisibility, paired with our work, reminds us of a pragmatic consideration: that we \textit{cannot} measure all of labor, certainly not to the desired level of quality. ``No representation of the world is either complete or permanent'' \cite[][p.257]{Gerson1986Analyzing} and even the act of ``making visible can incur invisibilities'' \cite[][p.25]{Star1999Layers}. Even with improved qualitative and quantitative measures (as we too offer in this work), it may not be meaningful or desirable to document some tasks to ensure privacy, autonomy, and flexibility. Emotion regulation, for example, may not be helpful to make record of in some contexts (e.g. deliberation over design changes), though it certainly may be in others (e.g. social work, therapy, ecosystem moderation).\footnote{Our thanks to an anonymous reviewer for motivating this point.}

With an eye toward equitable, balanced, and secure specifications for ``What should and shouldn't be visible?'' \cite{Star1999Layers}, responsive design strategies probably need participatory techniques for understanding ecosystems \cite{Germonprez2017Theory, Frey2019This}. A balanced design may therefore include participatory determination between representatives of an ecosystem's different stakeholder groups over what is meaningful to make visible (tasks, durations, difficulties, expertises, authorship, etc.); how to make them visible (recordkeeping, sharing, and accessibility designs); what to make invisible; and how to keep it invisible. This would help ensure the visibility of mutually-valued information while balancing a group's collective interests (motivations, security, privacy, information overload, etc.). Designing such a process for every ecosystem could be quite time consuming and ineffective at balancing stakeholder voices, though, so it may help for researchers to develop ``playbooks'' to ease (but not necessarily standardize) these efforts by helping stakeholders think through desired quantities and qualities of labor data to make visible and invisible \cite[c.f.][]{Google2024Attributing}. This might be aided by designing sufficient flexibility into collaborative platforms such that specific ecosystems can create contextually-defined artifacts for monitoring qualities of interest \cite{Meluso2022Flexible}.

%%% Implications --- Visibility Anchoring
However, our finding on visibility anchoring implies that OSS participants may assume that work is more visible than it actually is because ``openness seems like visibility'' to many. While a common assumption, this is not the case because visibility requires information to be recorded, shared, and accessible \cite{Stohl2016Managing} while openness typically falls along dimensions of transparency-opacity (roughly synonymous with visibility), inclusion-exclusion, and concentrated-distributed decision rights \cite{Splitter2023Openness}. So if participants presume that openness creates visibility, it may actually \textit{bias} participants into believing that invisible labor is less common than it really is, thereby decreasing attention, legitimacy, and compensation efforts. Our finding that visibility anchoring reduced credit importance perceptions reinforces this likelihood. Instead, designing ecosystems around an assumption that most information is \textit{not visible} --- despite being labeled ``open'' --- may be more prudent, particularly because it would incentivize stakeholders to think carefully about what they care enough about to make visible (and invisible) during participatory design processes.

Quantitatively and qualitatively, that an average 50\% of OSS labor gets compensated holds significant implications, too. Specifically, reducing invisibility requires greater compensation (in credit and pay). Again, while not all work should be fully visible, we join others who take the feminist position that workers should receive appropriate compensation for their work \cite{DIgnazio2020Data}, if an individual is so motivated (whether extrinsic, intrinsic, or internalized). If individuals have sufficient means (e.g. wealth, social status), they may be able to compensate themselves for performing invisible tasks, but most people cannot. We see this already in OSS when privileged individuals receive credit for activities that surpass community-defined thresholds of social status or centrality while less privileged individuals do not receive such credit \cite{DIgnazio2020Data, Young2021Which}. But OSS is too important to society to ignore that participants are regularly undercompensated, even in expensive forms (like attribution). Per our qualitative results, then, compensation may involve some combination of expressive attribution, instrumental attribution, non-attribution, pay, opportunities, etc. How individuals are compensated might also benefit from participatory design, this time in the context of an ecosystem's pragmatic constraints (e.g. individuals' evolving time commitments, individual and collective funding availability). Together, these efforts may help increase participant satisfaction with credit mediums (or attribution types) and frequencies by addressing participant motivations and airing to more stakeholders the pragmatic constraints on the ecosystem.

\subsection{Limitations \& Future Work}
\label{subsec:discussion_limitations}

%%% Limitations %%%
Our work has several limitations. First of all, survey instruments tend to prove more reliable in the form of validated scales. Questions remain about how feasible creating a scale would be for something so difficult to trace which motivated the exploratory nature of this study. That said, existing information visibility scales \cite{TerHoeven2021Assessing} may guide the labor visibility portion, leaving only need for compensation measures. Next, our sample sizes are too small to make these findings conclusive about all of OSS. In particular, we did not see differences between communities as we might have expected; breaking down our responses by artifacts and use cases did not show statistical differences between category means on any of the quantitative questions (0 of 16, 0\%). Sampling OSS ecosystems broadly as we did here may prove less helpful going forward than conducting targeted studies to benefit particular communities or study specific ecosystem characteristics, though. We were also limited in the questions we could ask participants because our survey spanned multiple ethnics regulatory jurisdictions. So future efforts might focus on specific artifacts, use cases, geographic regions, programming language communities, sizes of projects and ecosystems, volunteer statuses, corporate participation, years of participation, membership status, regulatory jurisdictions, and many other variables that were difficult to measure through an anonymous, self-reported survey. Third, our sample of convenience via social media prevented us from engaging with those who use social media less frequently meaning our sample may contain biases toward social media users and more socially-engaged ecosystems.

%%% Future Work %%%
Perhaps the greatest need arising from this work is a granular understanding of how technology affordances, co-evolving technologies and goals \cite{Leonardi2011When, Gibson2022Sustaining}, shape visibility and invisibility in OSS ecosystems. How do specific artifacts facilitate the existence of labor data, its sharing, accessibility, and participant compensation? And how do participants' motivations shift as a product of the capabilities of the resulting artifact? For example, we saw that certain attribution mediums corresponded to greater credit satisfaction. This was further supported by our qualitative analysis wherein participants preferred more or less automated and personalized attribution depending on their motivations. Further research might examine how interaction between credit mediums and motivations shape worker satisfaction (though other factors may influence satisfaction as our findings only show a correlation). These artifacts also shape worker perceptions of visibility, so researchers and practitioners likely need to pay careful attention to the interplay between accessibility and perceptions of openness. Labor will not become visible (to the desired level of stakeholders in each ecosystem) without ensuring consistent support for each factor that influences labor visibility and compensation. So, articulation work and infrastructuring activities would benefit from renewed attention, as would economic implications of whether or not invisible labor receives compensation.

Future work might also triangulate our findings within OSS through other qualitative and quantitative methods. We especially invite researchers to explore other factors that may influence invisible labor including technology affordances \cite{Leonardi2011When}, gender \cite{Frluckaj2022Gender, Berdahl2018Work}, governance \cite{Shah2006Motivation, Frey2019This}, and ecosystem size \cite{Geiger2021Labor}, each of which may have some effect. Studies of invisible labor need not limit themselves to OSS contexts either as invisible labor happens everywhere, in computing and more broadly. For CSCW, this article provides further evidence in support of what Toxtli \& colleagues \cite{Toxtli2021Quantifying} recently found in crowd work. We hope others will similarly build on our work in OSS and other domains.

%%%%%%%%%%%%%%%%%%%%%%%%%%%%%%%%%%%%%%%%%%%%%%%%%%%%%%%%%%%%%%%%%%%%%%%%%%%%%%%%%%%%%%%%%%%%%%%%%%%
\section{Conclusion}
\label{sec:conclusion}

% Summary
Much of the work that happens in OSS ecosystems is not visible, not compensated, or both. Such invisible labor makes it difficult for individuals to advance their careers, delegitimizies ``background work,'' and limits organizational understanding of crucial processes. In this study, we measured invisible labor in OSS ecosystems by estimating the visibility and compensation rates of OSS activities. We found that approximately half of OSS labor does not receive credit, a low-cost form of attribution that often precedes other forms of compensation, and that conflicting motivations amidst attribution practices may be partially responsible. In addition, we demonstrated that anchoring labor to the concept of visibility may lead participants to overestimate the visibility of their work and decrease its importance to them. Designing attribution and compensation systems with diverse stakeholder groups --- while assuming that most OSS work is \textit{not} visible or compensated --- may therefore help OSS ecosystems better satisfy the varied motivations of their participants.

\begin{acks}
Thanks to Jean-Gabriel Young, Melania Lavric, and the anonymous reviewers for their suggestions throughout. The authors were partially supported by Google under the Open Source Complex Ecosystems and Networks project. J.M. was partially supported by the Sloan Foundation through the Vermont Research Open Source Program Office, and by the NSF award BCS-2332598. M.Z.T. was partially supported by the Northeastern University Future Faculty Postdoctoral Fellowship Program.
\end{acks}


\appendix
\section{Invisible Labor Survey}
\label{sec:appendix_survey}

The following are the complete survey questions as constructed in Qualtrics. A consent form (Q1) preceded these questions with the title ``Open Source Credit Study'' and an invitation to participate in a research study ``about how people receive credit for the tasks they do as part of open source projects.'' \textit{Italicized text} indicates a question response constraint (e.g. a multiple choice question versus a multiple select question).

\begin{enumerate}[label={Q\arabic*.},font={\bfseries}]
    \item Consent form \textit{(Click to consent)}
    \item The questions in this survey ask about your experiences with \textbf{open source projects}. Specifically, the questions ask about projects you worked on \textbf{during the last 2 years}, and that \textbf{at least one other person also worked on}. \textit{(no response requested)}
    \item The following categories describe things that projects either create or work on. Which categories apply to the projects you worked on? Select all that apply. \textit{(multiple select)}
    \begin{itemize}
        \item Hardware
        \item Operating systems
        \item Languages
        \item Frameworks
        \item Libraries
        \item Packages
        \item Applications
        \item Data science
        \item Scripts
        \item Web-related
        \item Something else \textit{(with text field)}
    \end{itemize}
    \item Projects create for different use cases. Which cases apply to the projects you worked on? Select all that apply. \textit{(multiple select)}
    \begin{itemize}
        \item Browsers
        \item Business \& productivity applications
        \item Communication technologies
        \item Content management
        \item Education
        \item Entertainment
        \item Manufacturing
        \item Publishing
        \item Scientific research
        \item Software infrastructure
        \item Something else \textit{(with text field)}
    \end{itemize}
    \item Tasks vary in shape, size, and how visible they are to other people. For about 30 seconds, \textbf{think through the different kinds of tasks you have done} as part of these projects over the past 2 years. When you are ready, click on the right arrow to go on to the next question. \textit{(no response requested)}
    \item The following questions ask about receiving credit. When we talk about having \textbf{received credit for a task}, we mean that the project told someone that you did something for the project. Those people may also work on the project or may be outside of the project. \textit{(no response requested)}
    \item In the past two years, from how many projects did you receive credit for your tasks? \textit{(multiple choice)}
    \begin{itemize}
        \item 1 - None of them
        \item 2 - Few of them
        \item 3 - Some of them
        \item 4 - Most of them
        \item 5 - All of them
        \item I'm not sure
    \end{itemize}
    \item In the past two years, for how many of your tasks did you receive credit? \textit{(multiple choice)}
    \begin{itemize}
        \item 1 - None of them
        \item 2 - Few of them
        \item 3 - Some of them
        \item 4 - Most of them
        \item 5 - All of them
        \item I'm not sure
    \end{itemize}
    \item Projects give people credit for tasks through different mediums. Through what mediums did you receive credit? Select all that apply. \textit{(multiple select)}
    \begin{itemize}
        \item Automated metrics (e.g. GitHub Contributors)
        \item Project membership (e.g. GitHub Organizations)
        \item Project documentation (e.g. acknowledgement lists, release notes)
        \item Standardized tools (e.g. All Contributors, CRediT)
        \item Social media (e.g. Twitter, Facebook)
        \item Blogs (e.g. Medium, personal websites)
        \item Presentations (e.g. talks, conferences)
        \item Something else \textit{(with text field)}
        \item I'm not sure
    \end{itemize}
    \item How satisfied are you with \textbf{the mediums} through which you receive credit? \textit{(multiple choice)}
    \begin{itemize}
        \item 1 - Extremely dissatisfied
        \item 2 - Dissatisfied
        \item 3 - Neither satisfied nor dissatisfied
        \item 4 - Satisfied
        \item 5 - Extremely satisfied
    \end{itemize}
    \item Think about all the tasks you performed again. \textit{(no response requested)}
    \item \textit{(Randomized)} How often did \textbf{2 or more people} know that you performed those tasks? \textit{(multiple choice)}
    \begin{itemize}
        \item 1 - Never
        \item 2 - Rarely
        \item 3 - Sometimes
        \item 4 - Often
        \item 5 - Always
        \item I'm not sure
    \end{itemize}
    \item How often did \textbf{1 other person} know that you performed those tasks? \textit{(multiple choice)}
    \begin{itemize}
        \item 1 - Never
        \item 2 - Rarely
        \item 3 - Sometimes
        \item 4 - Often
        \item 5 - Always
        \item I'm not sure
    \end{itemize}
    \item \textit{(Randomized)} How often did \textbf{nobody else} know that you performed those tasks? \textit{(multiple choice)}
    \begin{itemize}
        \item 1 - Never
        \item 2 - Rarely
        \item 3 - Sometimes
        \item 4 - Often
        \item 5 - Always
        \item I'm not sure
    \end{itemize}
    \item How satisfied are you with \textbf{how many} of your tasks received credit? \textit{(multiple choice)}
    \begin{itemize}
        \item 1 - Extremely dissatisfied
        \item 2 - Dissatisfied
        \item 3 - Neither satisfied nor dissatisfied
        \item 4 - Satisfied
        \item 5 - Extremely satisfied
    \end{itemize}
    \item How important is it to you to receive credit for the tasks you do? \textit{(multiple choice)}
    \begin{itemize}
        \item 1 - Not at all important
        \item 2 - Slightly important
        \item 3 - Moderately important
        \item 4 - Very important
        \item 5 - Extremely important
    \end{itemize}
    \item Think about how the projects you worked on gave people credit. From your perspective, \textbf{what worked well} about how they gave people credit? \textit{(free response field)}
    \item \textbf{What did not work well} about how they gave people credit? \textit{(free response field)}
    \item Is there anything else that you’d like to share with us? \textit{(free response field)}
\end{enumerate}

\end{document}